\documentclass[aps,prl,reprint,superscriptaddress,nofootinbib,amsmath,amssymb]{revtex4-2}
\usepackage[T1]{fontenc}
\usepackage{mathptmx,graphicx,booktabs,longtable,microtype,bm}
\usepackage[unicode=true,pdfencoding=auto,colorlinks=true,allcolors=blue,hypertexnames=false]{hyperref}
\newcommand{\etal}{\textit{et al.}}
\begin{document}

\title{Nuclear Precision and Multimessenger Constraints on Neutron-Star Structure}
\author{Hisham Anwer}
\affiliation{Physics of Universe Program, Zewail City of Science and Technology, Giza, Egypt}
\affiliation{Physics Department, Faculty of Science, Cairo University, 12613 Giza, Egypt}
\author{A. Adel}
\affiliation{Physics Department, Faculty of Science, Cairo University, 12613 Giza, Egypt}
\author{E. Omar}
\affiliation{Physics of Universe Program, Zewail City of Science and Technology, Giza, Egypt}

\begin{abstract}
Neutron-star cores retain structural freedom beyond the constraints supplied by nuclear measurements. We quantify this freedom through an observable-dependent variance ceiling and identify radius--Love compensation as a mechanism suppressing nuclear constraints on tides. Across four data-conditioned equation-of-state families, a pulsar radius measured to 0.3 km paired with a binary tide measured to 10\% removes 44--68\% of the current logarithmic Love-number variance. The framework connects laboratory precision and complementary stellar measurements to core structure.
\end{abstract}

\makeatletter
\let\arxivOriginalMaketitle\maketitle
\def\maketitle{%
  \@author@finish
  \global\let\arxivAuthors\@AAC@list
  \global\let\arxivAffiliations\@AFF@list
  \global\let\arxivAuthorGroups\@AFG@list
  \arxivOriginalMaketitle
}
\makeatother
\maketitle

Neutron skins connect laboratory measurements of finite nuclei to the properties of neutron-rich matter in neutron stars. Correlations among skins extend information from measured lead to heavy and superheavy nuclei, while their shared sensitivity to the isovector interaction links these observables to the pressure of neutron-rich matter \cite{brown,centelles,lattimer04,oertel17,anwerprd}. The resulting constraints on stellar structure depend on both the nuclear interaction and the high-density equation of state (EOS). Distinct high-density EOSs can share identical nuclear inputs yet predict different stellar radii and density profiles. Quantifying how improved nuclear precision constrains each stellar observable therefore identifies the information accessible to laboratory measurements and the additional information supplied by complementary measurements of neutron-star radii and tidal deformabilities.

Laboratory measurements and neutron-star observations constrain complementary parts of this connection. Parity-violating electron scattering measures neutron skins in $^{208}$Pb (PREX-II) and $^{48}$Ca (CREX) \cite{prex,crex}. Gravitational waves (GW) and pulse profiles from the Neutron Star Interior Composition Explorer (NICER) constrain tidal deformation and stellar radii \cite{gw,gwprops,riley19,salmi24,miller26}. Existing studies connect nuclear observables to stellar properties and merger emission while establishing the influence of high-density modeling \cite{reed21,essick21,legred22,legred24,anwermerger26}; prospective analyses quantify the return of improved stellar precision \cite{li26}. Quantifying the return of improved nuclear precision requires resolving both the shared nuclear information and the core variation at fixed nuclear input.

We quantify an observable-dependent ceiling on nuclear experimental return and identify conditional radius--Love compensation near $1.4M_\odot$, where $M_\odot$ is the solar mass. Three nuclear calibrations and two core constructions test the radius-over-tide ordering. Pairing a pulsar radius with a binary tide then accesses density-profile information left unresolved by nuclear precision.

\textit{Nuclear information and high-density uncertainty.}
A lead-skin measurement probes the shared isovector structure of nuclei. The skin $\Delta=R_n-R_p$ is the difference between point-neutron and point-proton root-mean-square radii; PREX-II gives $\Delta(^{208}\mathrm{Pb})=0.283\pm0.071$ fm \cite{prex}. Isotopic-chain mappings describe skin evolution with neutron number. The stellar calculation propagates the nuclear calibration uncertainty for the fixed $^{208}$Pb observable into nuclear matter; a subsequent lead measurement reduces this uncertainty according to alignment and calibration width \cite{reinhard10,hu22,unedf0,unedf1}. We denote the nuclear coordinates by $\theta$ and retain their correlations when generating matter in cold beta equilibrium with electrons and muons.

High-density EOSs complete the stellar model beyond the laboratory anchor. At matching density $n_m=1.1n_0$, $1.25n_0$ or $1.5n_0$, where $n_0$ is nuclear saturation density, the primary family attaches causal two-segment cores with optional energy-density jumps \cite{css}. Physical stability and maximum mass $\ge2M_\odot$ select 6,885 EOS paths from an empirical UNEDF0 calibration prior, retaining 224 distinct nuclear draws \cite{supp}. Stellar likelihoods combine GW170817, J0030, updated J0740 and J0437 data, retaining source-prior and stable-branch support \cite{gwprops,riley19,salmi24,miller26,supp}. UNEDF0 and UNEDF1 calibrations, a distinct covariant nuclear posterior and a continuous sound-speed construction provide additional tests below. The present stellar posteriors combine nuclear calibrations with GW and NICER data; treating PREX-II and CREX separately isolates the return of an additional laboratory measurement. These stellar posteriors provide the baseline for the independent prospective 0.030-fm lead experiment, before PREX-II or CREX conditioning.

Residual high-density uncertainty sets an upper bound on the variance reduction achievable through nuclear measurements. At fixed nuclear input, this is the spread across allowed cores, matching densities and configurations. Let $\mathcal D$ denote present data and $\phi$ the core, matching and configuration variables. For a stellar observable $X$, the conditional mean $\mu_X(\theta)=\mathbb E(X|\theta,\mathcal D)$ averages the allowed cores, matches and configurations at fixed nuclear input. With $V_X=\operatorname{Var}(X|\mathcal D)$, total variance gives
\begin{align}
 V_X&=\mathcal B_{{\rm core},X}+\operatorname{Var}_{\theta|\mathcal D}\mu_X,\notag\\
 \mathcal B_{{\rm core},X}&=\mathbb E_{\theta|\mathcal D}\operatorname{Var}(X|\theta,\mathcal D).
 \label{eq:budget}
\end{align}
A nuclear-only measurement $Y$ satisfies $Y\perp\phi\mid\theta,\mathcal D$. Averaging over its possible outcomes yields
\begin{equation}
0\le\mathcal T_X\equiv1-\frac{\mathbb E_Y\operatorname{Var}(X|\mathcal D,Y)}{V_X}
\le1-\frac{\mathcal B_{{\rm core},X}}{V_X}.
\label{eq:bound}
\end{equation}
The ceiling identifies the stellar variance accessible through nuclear coordinates, while $\mathcal B_{{\rm core},X}$ measures the complementary core contribution. A specific experiment recovers the part of $\mu_X$ resolved by its outcomes. Conditional independence and total variance give the exact identity
\begin{equation}
 \mathcal T_X=\frac{\operatorname{Var}_{Y|\mathcal D}\!\left[\mathbb E(\mu_X|\mathcal D,Y)\right]}{V_X}.
 \label{eq:recovered}
\end{equation}
Thus laboratory reach combines the available nuclear variance with the experiment's ability to distinguish nuclear inputs having different stellar conditional means. Exact nuclear coordinates attain the ceiling; finite precision and sensitivity to particular nuclear directions determine the realized gain. The ceiling is the nuclear-group variance-based sensitivity index, or normalized partial perfect-information value under squared-error loss \cite{oakley04}. Equations~\eqref{eq:bound} and \eqref{eq:recovered} connect this established variance decomposition to observable-specific neutron-star forecasts.

\begin{figure}[t]\centering
\includegraphics[width=\columnwidth]{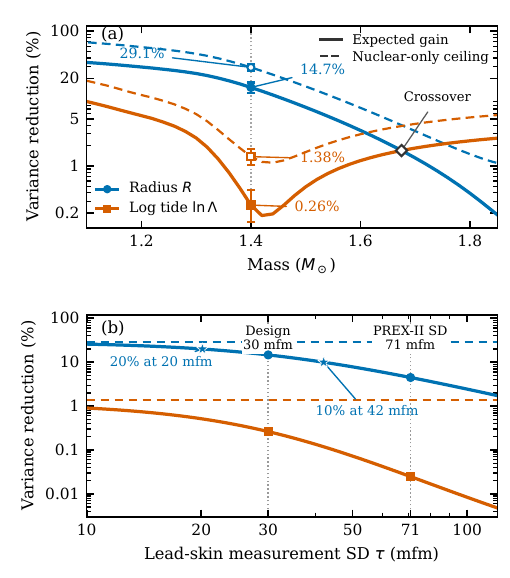}
\caption{Nuclear reach in the primary combined posterior. Solid curves give the expected independent lead-measurement gain; dashed curves give nuclear-only ceilings. (a) Mass dependence at SD 0.030 fm. Filled/open markers denote gains/ceilings with 95\% whole-draw sensitivity intervals. The diamond marks the ordering crossover. (b) Precision dependence at $1.4M_\odot$; stars mark 10\% and 20\% radius gains. Vertical lines mark the 30-mfm design and PREX-II's 71-mfm SD as precision comparators, without conditioning on PREX-II's central value. Here $1\,\mathrm{mfm}=10^{-3}$ fm. Vertical axes are logarithmic; each observable uses its own current variance.}\label{fig:transmission}
\end{figure}

\begin{figure}[t]\centering
\includegraphics[width=\columnwidth]{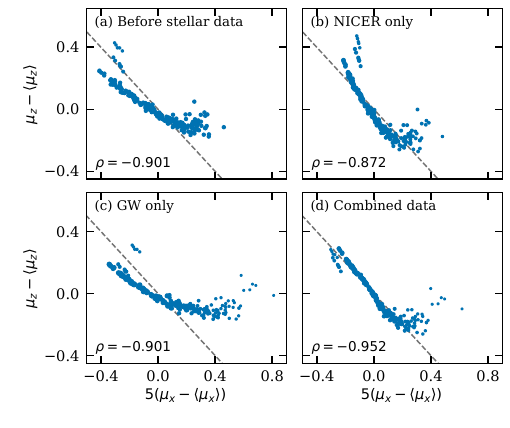}
\caption{Observations reshape radius--Love compensation in the primary UNEDF0 family at $1.4M_\odot$. For $x=\ln(R/1\,\mathrm{km})$ and $z=\ln k_2$, each point is a conditional mean over cores, matches and branches at one nuclear draw; marker area increases with its probability. Brackets denote the nuclear-draw average. Centered coordinates sum to the conditional change in $\ln\Lambda$. The dashed line gives the theoretical cancellation condition $5(\mu_x-\langle\mu_x\rangle)+(\mu_z-\langle\mu_z\rangle)=0$. All panels share axes.}\label{fig:compensation}
\end{figure}

Radii and tides retain different amounts of nuclear information (Fig.~\ref{fig:transmission}). At $1.4M_\odot$, an independent lead-skin measurement with standard deviation (SD) 0.030 fm gives a 14.7\% expected reduction of radius variance but 0.26\% of $\ln\Lambda$ variance. Their nuclear-only ceilings are 29.1\% and 1.38\%. Radius gains at 1.2 and $1.6M_\odot$ are 29.5\% and 3.4\%. The ordering reverses near $1.7M_\odot$ in this posterior, where high-density uncertainty leaves little constraining power for radius. Comparing $\ln R$ with $\ln\Lambda$ preserves the hierarchy (Table~\ref{tab:hierarchy}). At fixed matching density the log-tidal ceilings span 3.9--18.1\%. Marginalizing the match allows its conditional weights to depend on the nuclear draw; the pooled ceiling can lie below every fixed-match value. The ceiling quantifies the nuclear information available within the conditioned EOS family. The 0.030-fm benchmark is close to the projected Mainz Radius Experiment (MREX) goal of 0.5\% in the lead neutron radius \cite{mrex23}. In the primary combined posterior, lead SDs of 0.042 and 0.020 fm yield 10\% and 20\% radius-variance reductions at $1.4M_\odot$ \cite{supp}.

Data-conditioned radius--Love compensation accounts for the primary posterior's strongest suppression of nuclear constraining power on tides. At fixed mass, the radius factor increases the tide, while $k_2$ measures the normalized quadrupolar response of the internal mass distribution. Opposing conditional changes nearly cancel when their amplitudes match. The quadrupolar deformability is $\Lambda=(2/3)k_2C^{-5}$, where $k_2$ is the Love number, $C=GM/(Rc^2)$ is compactness, $G$ is Newton's constant and $c$ the speed of light \cite{hinderer}. At fixed mass set $x=\ln(R/1\,\mathrm{km})$, $z=\ln k_2$ and $\ell=\ln\Lambda$. Their conditional means obey $\mu_\ell=5\mu_x+\mu_z+\mathrm{constant}$, giving
\begin{align}
 \operatorname{Var}_\theta\mu_\ell={}&25\operatorname{Var}_\theta\mu_x+\operatorname{Var}_\theta\mu_z\notag\\
 &+10\operatorname{Cov}_\theta(\mu_x,\mu_z).
 \label{eq:compensation}
\end{align}
At $1.4M_\odot$ in the primary posterior, contributions of 14.08\%, 14.36\% and $-27.07\%$ of total log-tidal variance leave 1.38\%, also obtained by evaluating $\operatorname{Var}_\theta(5\mu_x+\mu_z)$ directly. The conditional-mean correlation is $-0.952$, whereas the full paired stellar sample has positive covariance. The cancellation removes 95.2\% of the sum of the two positive between-draw terms. Small residual variance requires both opposing conditional shifts and comparable amplitudes. For $u=5\mu_x$, $v=\mu_z$, $V_u=\operatorname{Var}_\theta u>0$, $\nu=\sqrt{\operatorname{Var}_\theta v/V_u}$ and $\rho=\operatorname{Corr}_\theta(u,v)$,
\begin{equation}
 \operatorname{Var}_\theta\mu_\ell=V_u\big[(\nu-1)^2+2\nu(1+\rho)\big].
 \label{eq:amplitudes}
\end{equation}
The two nonnegative terms separate amplitude mismatch from imperfect anticorrelation. Figure~\ref{fig:compensation} resolves the observational updates; a distinct microscopic framework tests amplitude matching below. Reconstructed profiles reveal the structural origin of this compensation in the primary ensemble. Larger conditional radii accompany a more extended outer region and a more centrally concentrated mass distribution, reducing the normalized tidal response $k_2$ and offsetting the geometric radius contribution \cite{supp}.

\begin{table}[t]\centering\setlength{\tabcolsep}{3pt}
\caption{Laboratory reach at $1.4M_\odot$. Gains are outcome-averaged variance reductions from an independent 0.030-fm lead measurement. Ceilings apply to any nuclear-only experiment within the stated family. Entries are percentages; $x=\ln(R/1\,\mathrm{km})$, $\ell=\ln\Lambda$. Matching densities are pooled.}\label{tab:hierarchy}
\begin{ruledtabular}\begin{tabular}{lrrrrr}
 & \multicolumn{3}{c}{Lead gain} & \multicolumn{2}{c}{Ceiling}\\
Family / update & $R$ & $x$ & $\ell$ & $R$ & $\ell$\\
\hline\multicolumn{6}{c}{UNEDF0, two-segment cores}\\\hline
Before stellar data & 20.33 & 19.67 & 3.23 & 33.89 & 5.73\\
NICER only & 17.00 & 16.59 & 1.16 & 29.16 & 4.81\\
GW only & 16.89 & 16.57 & 3.45 & 33.53 & 6.98\\
Combined data & 14.68 & 14.38 & 0.26 & 29.11 & 1.38\\
\hline\multicolumn{6}{c}{Continuous sound speed, combined data}\\\hline
UNEDF0 & 25.72 & 25.40 & 2.40 & 50.61 & 5.58\\
UNEDF1 & 5.43 & 5.42 & 0.47 & 23.92 & 3.19\\
Covariant EDF & 1.35 & 1.35 & 0.55 & 4.46 & 1.85\\
\end{tabular}\end{ruledtabular}\par\smallskip{\raggedright\footnotesize Covariant gains have 95\% whole-draw sensitivity intervals of 0.43--2.71\% ($R$) and 0.26--0.95\% ($\ell$) \cite{supp}.\par}\end{table}

\begin{figure}[t]\centering
\includegraphics[width=\columnwidth]{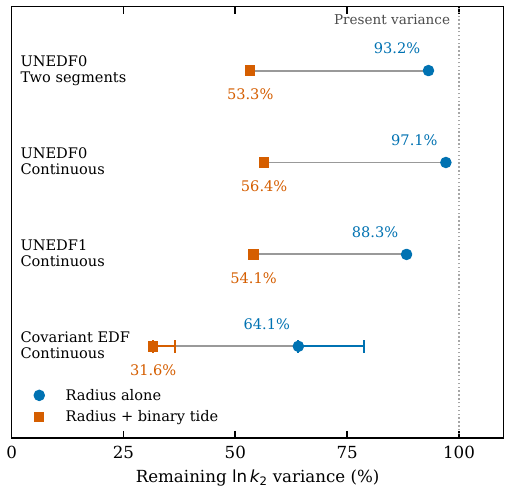}
\caption{Complementary observations access density-profile information across four updated families. Points give the expected remaining $\ln k_2(1.4M_\odot)$ variance after a 0.3-km-SD pulsar radius alone or that radius paired with a different binary's effective tide at 0.1 log-tidal SD. Each row is normalized to its own current posterior variance; connecting lines guide the comparison. Source masses are specified in the text. All cases average over predictive outcomes and use independent measurement noise. Covariant-row whiskers span the nested and independent sampling controls; they are not credible intervals.}\label{fig:profile}
\end{figure}

Observational updates reveal how the compensation develops (Fig.~\ref{fig:compensation} and Table~\ref{tab:hierarchy}). In the primary family, the log-tidal lead gain is 3.23\% before stellar data, 1.16\% with NICER alone, 3.45\% with GW alone and 0.26\% jointly. NICER produces the standalone suppression, and the combined likelihood selects the strongest compensation. The update changes both the probabilities of nuclear draws and the core distribution within each draw. Whole-draw resampling gives 95\% sensitivity intervals of 12.1--17.4\% for the radius gain, 0.15--0.43\% for the log-tidal gain and 1.03--1.81\% for its ceiling \cite{supp}.

A continuous sound-speed family tests the transmission beyond the primary core construction. Its squared sound speed $s=dP/d\epsilon=c_s^2/c^2$, for pressure $P$, energy density $\epsilon$ and sound speed $c_s$, is anchored to the microscopic EOS and varied at three higher densities. Both UNEDF0 and UNEDF1 generate independent nonlinear microscopic matter and stellar solutions. The same observational likelihoods update 71,639 accepted paths across three matches \cite{supp}. The canonical-mass ordering persists for both radius and log radius across the four conditioned families (Table~\ref{tab:hierarchy}).

A covariant nuclear posterior tests the microscopic framework itself. We use paired parameter and neutron-skin samples from a nonlinear relativistic functional calibrated to finite-nucleus energies and charge radii \cite{giuliani23,chen14}. With the same core rule and matches, 40,562 EOS paths survive screening. The conditional correlation is -0.959, but $\nu=1.89$ differs from the primary value 1.01. This comparison isolates the amplitude requirement: unequal amplitudes leave a larger residual. Here GW alone gives a 0.12\% log-tidal gain, NICER alone 2.53\% and the joint update 0.55\%, exposing the framework dependence of source attribution. At $1.6M_\odot$, the gains become 0.19\% for radius and 1.85\% for log tide, revealing the mass dependence of nuclear constraining power \cite{supp}.

\textit{Accessing the surviving density-profile information.}
Complementary stellar observations can probe core variation left unresolved by a laboratory anchor. Approximate compactness--Love relations organize the dominant radius dependence \cite{yagiprd13,maselli13,legred24}, yet held-nuclear-draw validation leaves 30.9\% of log-tidal variance in a cubic radius predictor's error. Two accepted cores with identical nuclear input illustrate the freedom: radii separated by only 2.86 m near 13.025 km coexist with a 10.64\% difference in tide. This density-profile freedom motivates a complementary stellar measurement.

Complementary observations of separate stellar systems constrain their shared EOS. We pair a prospective radius measurement of a $1.418M_\odot$ pulsar with a binary tidal measurement at chirp mass $\mathcal M=1.186M_\odot$ and mass ratio $q=m_2/m_1=0.85$. Here $\mathcal M=(m_1m_2)^{3/5}/(m_1+m_2)^{1/5}$; the component masses are approximately $1.479M_\odot$ and $1.257M_\odot$. The binary measurement constrains the effective tidal deformability,
\begin{equation}
 \widetilde\Lambda=\frac{16[(1+12q)\Lambda_1+(q+12)q^4\Lambda_2]}{13(1+q)^5}.
\end{equation}
Here $\Lambda_1$ and $\Lambda_2$ are the dimensionless component tidal deformabilities. Averaging over supported branches at each EOS, we predict $z=\ln k_2$ for a reference $1.4M_\odot$ star. Independent Gaussian errors with radius SD 0.3 km and log-tidal SD 0.1 (approximately 10\%) define prospective targets for next-generation X-ray timing and third-generation GW observations, with source masses held fixed. The radius benchmark specifies precision on the nonrotating EOS radius; pulsar applications require a spin-dependent mapping \cite{supp}.

The incremental value of the binary tide has a direct variance interpretation. Let $Y_R$ and $Y_T$ denote the prospective radius and tidal measurements, and $V_z=\operatorname{Var}(z|\mathcal D)$. Their additional outcome-averaged gain obeys
\begin{equation}
 \mathcal G_{T|R}=\frac{\mathbb E_{Y_R|\mathcal D}\operatorname{Var}_{Y_T|\mathcal D,Y_R}
 [\mathbb E(z|\mathcal D,Y_R,Y_T)]}{V_z}\ge0.
 \label{eq:increment}
\end{equation}
Here $\mathcal G_{T|R}$ is the expected variance after radius alone minus that after both measurements, normalized by $V_z$. The tide resolves differences among EOSs compatible with the radius measurement, directly connecting the paired forecast to density-profile information.

A binary tide consistently adds information that a pulsar radius leaves largely unresolved. With independent Gaussian errors at these precisions, radius alone leaves 64.1--97.1\% of the current $z$ variance, whereas the paired observations leave 31.6--56.4\%, averaged over their possible outcomes (Fig.~\ref{fig:profile}).\ Microscopic and core priors set the starting $\ln k_2$ variances and the fractional returns. Covariant sampling controls leave 31.6--36.6\% with the pair, compared with 64.1--78.9\% for radius alone; predictive quadrature errors are much smaller \cite{supp}.

Our framework quantifies how nuclear experimental precision and complementary stellar observations constrain neutron-star core structure, identifying the information accessible to laboratory measurements and the density-profile information revealed by combining radius and tidal observations. Across four conditioned families, the canonical-mass radius-over-tide ordering persists with mass-dependent crossovers; paired observations remove 44--68\% of current $\ln k_2$ variance at the stated precisions. Nuclear measurements locate the shared low-density pressure direction, while paired radii and tides resolve the internal structure sustained by high-density matter. Together these channels provide a quantitative route from improved experimental precision to complementary constraints on neutron-star cores.

\textit{Data availability.} The data that support the findings of this study are available from the corresponding author upon reasonable request.
\begingroup\sloppy
\endgroup

\clearpage
\onecolumngrid
\setcounter{page}{1}
\setcounter{equation}{0}
\setcounter{figure}{0}
\setcounter{table}{0}
\setcounter{section}{0}
\setcounter{subsection}{0}
\renewcommand{\thepage}{\arabic{page}}
\renewcommand{\theequation}{S\arabic{equation}}
\renewcommand{\thetable}{S\Roman{table}}
\renewcommand{\thefigure}{S\arabic{figure}}
\setlength{\emergencystretch}{2em}
\setcounter{secnumdepth}{2}
\renewcommand{\thesection}{S\arabic{section}}
\renewcommand{\thesubsection}{\thesection.\Alph{subsection}}
\makeatletter
\renewcommand{\p@subsection}{}
\makeatother
\hypersetup{bookmarksnumbered=true}

\makeatletter
\begingroup
\let\@AAC@list\arxivAuthors
\let\@AFF@list\arxivAffiliations
\let\@AFG@list\arxivAuthorGroups
\def\@title{Supplemental Material: Nuclear Precision and Multimessenger Constraints on Neutron-Star Structure}
\let\@date\@empty
\let\frontmatter@abstract@produce\relax
\let\label\@gobble
\titleblock@produce
\endgroup
\def\endthebibliography{%
  \auto@bib@innerbib
  \edef\@currentlabel{\arabic{NAT@ctr}}%
  \label{SM:LastBibItem}%
  \endNAT@thebibliography
  \aftergroup\auto@bib@empty
}
\makeatother

\thispagestyle{article}

\section{Nuclear inputs and primary stellar construction}\label{SM:sec:inputs}
\subsection{Local covariance and configuration tracking}
A lead skin measurement constrains other nuclear observables through their shared sensitivity to the interaction parameters. We quantify this connection at fixed proton and neutron numbers, separating calibration uncertainty from isotope evolution. Let $\Delta=R_n-R_p$ denote the neutron skin, the difference between point neutron and point proton root mean square radii, expressed in femtometres (fm). Let $\theta$ denote the nuclear interaction coordinates. In the local Gaussian approximation, $\theta\sim\mathcal N(\theta_0,C_\theta)$, where $\theta_0$ is the calibration mean and $C_\theta$ its parameter covariance. For an observable $X$ of a fixed nucleus, $X_0=X(\theta_0)$ and $g_X=\nabla_\theta X|_{\theta_0}$ denote its baseline value and gradient. Linearization gives $X=X_0+g_X^T(\theta-\theta_0)$, with superscript $T$ denoting transpose. Within this subsection $P$ denotes the lead skin, with $g_P=\nabla_\theta P|_{\theta_0}$. Elsewhere $P$ denotes pressure. Define $v_P=g_P^TC_\theta g_P$, $v_X=g_X^TC_\theta g_X$ and $c_{XP}=g_X^TC_\theta g_P$. The operators $\mathbb E$ and $\operatorname{Var}$ denote expectation and variance. Completing the square for a Gaussian measurement $d_P$ with noise variance $\tau_P^2$ gives
\begin{align}
 \mathbb E(X|d_P)&=X_0+\frac{c_{XP}}{v_P+\tau_P^2}(d_P-P_0),\\
 v_{X|d_P}&=v_X-\frac{c_{XP}^2}{v_P+\tau_P^2},\\
 \eta_X&=1-v_{X|d_P}/v_X
 =\rho_{XP}^2\frac{v_P}{v_P+\tau_P^2}.
\end{align}
Here $P_0=P(\theta_0)$ is the calculated lead mean, $v_{X|d_P}=\operatorname{Var}(X|d_P)$ is the conditioned variance, and $\rho_{XP}=c_{XP}/\sqrt{v_Xv_P}$ the alignment. The corresponding fractional reduction in standard deviation (SD) is $1-\sqrt{1-\eta_X}$. The quantity $\eta_X$ measures variance reduction.

The independent nuclear calibrations give different prior widths even when their sensitivity directions align. Table~\ref{SM:tab:local} compares results for proton number $Z=120$ and neutron number $N=184$. UNEDF0's published and rescaled local widths use the same matrix compatible with the rounded correlations. These local Gaussian diagnostics are distinct from the empirical nuclear prior defining the nonlinear stellar budgets. UNEDF1 uses its own active coordinate basis and original covariance \cite{SM:unedf0,SM:unedf1,SM:haverinen,SM:hfbthoad}. Its lead SD is 0.01870 fm. The published UNEDF0 lead SD 0.0579 fm is distinct from the superheavy SD 0.0592 fm. These SDs refer to different observables evaluated at fixed nuclei. The local diagnostic uses the lead measurement from parity violating electron scattering in $^{208}$Pb (PREX-II), while CREX is the corresponding experiment on $^{48}$Ca \cite{SM:prex,SM:crex}. The stellar posteriors instead combine the original nuclear calibrations with gravitational wave (GW) data and pulse profiles from the Neutron Star Interior Composition Explorer (NICER), before conditioning on PREX-II or CREX. The prospective stellar gains average over outcomes of a new independent lead measurement with SD 0.030 fm.
\begin{table}[ht]\centering
\caption{PREX-II conditioning of the $Z=120,N=184$ skin before nonlinear stability selection. Both SD columns describe this superheavy skin, in fm. The conditioned SD uses the measured lead skin $0.283\pm0.071$ fm. Gain is the removed fraction of calibration variance.}\label{SM:tab:local}
\begin{tabular}{llrrr}\toprule
Interaction & Width convention & Prior SD & Conditioned SD & Gain (\%)\\\midrule
UNEDF0 & Published & 0.0592 & 0.0459 &39.9\\
UNEDF0 & Rescaled (local) & 0.0496 & 0.0409 &31.8\\
UNEDF1 & Original & 0.0183 & 0.0177 &6.5\\\bottomrule\end{tabular}\end{table}

Gradients follow tracked axial Hartree--Fock--Bogolyubov (HFB) configurations, evaluated with HFBTHO \cite{SM:hfbtho3}. Independent native/custom UNEDF1 baselines agree at printed precision. Halving the symmetry energy and density slope parameter steps changes gradients by less than 0.9\%. The largest difference scaled by the parameter SD is 0.083 millifemtometres (mfm), where $1\,\mathrm{mfm}=10^{-3}$ fm. Ten matched UNEDF0 configuration comparisons retain the tracked ordering: nine competing states are 1.5 to 8.1 MeV higher, and the negative three quarter SD $N=198$ comparison leaves the prolate state lower by 1.320 MeV. The negative full SD tracked attempt did not converge and supplies no branch ordering result. The same parameter point also fails the uniform matter stability screen at all three matches, so it is absent from the stellar family for an independent physical reason. These checks support local tracking on tested configurations, not global ground state certification.

\subsection{Microscopic matter and stellar structure}
The calibrated interaction determines the pressure and composition of matter below the core matching density. Let $n_0$ denote nuclear saturation density. We evaluate the full Skyrme expression for cold matter in beta equilibrium with electrons and muons. Write $n=n_n+n_p$ for baryon density and $\delta=(n_n-n_p)/n$ for neutron excess. The symmetry energy is $S(n)=\tfrac12\partial^2 e(n,\delta)/\partial\delta^2|_{\delta=0}$ for baryonic energy per particle $e$, and $L=3n_0S'(n_0)$ is its saturation slope. Surface, pairing and spin orbit terms vanish in uniform matter, but their calibration correlations with bulk parameters remain in $C_\theta$.

At matching densities $n_m=1.1n_0$, $1.25n_0$ or $1.5n_0$, the microscopic equation of state (EOS) joins a high density core. The matching pressure and energy density are $P_m$ and $\epsilon_m$. Each core segment satisfies $dP/d\epsilon=s_j$, with $c$ the speed of light, $c_{s,j}$ the segment sound speed and $s_j=c_{s,j}^2/c^2$ its square in units of $c^2$, pressure $P$ and total energy density $\epsilon$. Its EOS is
\begin{equation}
 \epsilon(P)=\epsilon_j+(P-P_j)/s_j.
\end{equation}
The first segment has $s_1=0.2,0.6,1$, and the second $s_2=1/3,1$. The transition lies 150 MeV fm$^{-3}$ above the matching energy density on its lower side, and the jump is $\Delta\epsilon=0$ or 150 MeV fm$^{-3}$. Thus $P_t=P_m+s_1(150\,\mathrm{MeV\,fm}^{-3})$. Subscripts $m,t,j$ denote the microscopic match, core transition and segment lower boundary. Here $n_j$ is baryon density at the segment boundary. Density follows the zero temperature first law,
\begin{equation}
 n/n_j=[(\epsilon+P)/(\epsilon_j+P_j)]^{1/(1+s_j)}.
\end{equation}
At a discontinuity, equality of pressure and chemical potential fixes the density jump. A thermodynamically matched polytropic crust joins the stable microscopic segment at its upper spinodal boundary. The grid follows the constant sound speed construction of Ref.~\cite{SM:css}.

Uniform matter and stellar structure are evaluated nonlinearly, while UNEDF lead skins use the local HFB sensitivities validated above. The UNEDF0 stellar integration is defined by the fixed empirical nuclear prior $\pi_0=256^{-1}\sum_{i=1}^{256}\delta_{\theta_i}$, where $\theta_i$ are the ten parameter nuclear nodes and $\delta_{\theta_i}$ is unit point mass at node $i$. Each node has equal weight before microscopic and stellar screening. The pre screening ensemble has lead skin SD 0.04738 fm. Subsequent physical acceptance and stellar likelihoods determine the working nuclear weights. The published width local covariance diagnostic above is a separate calculation.

The three matches and twelve cores per match initially have equal probability. A Sobol sequence transformed to a correlated Gaussian samples 256 nuclear parameter draws in normalized UNEDF0. Physical screens require positive symmetry energy, stable causal uniform matter up to the match, and maximum stellar mass $M_{\max}\ge2M_\odot$, where $M_\odot$ is the solar mass. The 6,885 surviving paths contain 224 distinct draws. Global normalization preserves the different acceptance fractions among nuclear draws and retains their relative statistical weights. The primary multimessenger update uses UNEDF0. The continuous core families below are evaluated before and after the same observational likelihoods.

Nonrotating stellar structure and tidal deformability follow the Tolman--Oppenheimer--Volkoff (TOV) equations and quadrupolar perturbation equations \cite{SM:hinderer}. With Newton's constant $G$ and the speed of light $c$ set to unity, the equations for enclosed gravitational mass $m(r)$ and pressure at areal radius $r$ are
\begin{equation}
 \frac{dm}{dr}=4\pi r^2\epsilon,\qquad
 \frac{dP}{dr}=-\frac{(\epsilon+P)(m+4\pi r^3P)}{r(r-2m)}.
\end{equation}
At a Maxwell interface with enclosed mass $m_t$, radius $r_t$ and pressure $P_t$, the usual logarithmic perturbation variable $y=rH'/H$, for quadrupolar metric perturbation $H$, with prime denoting $d/dr$, obeys
\begin{equation}
 y_{\rm out}=y_{\rm in}-\frac{4\pi r_t^3\Delta\epsilon}{m_t+4\pi r_t^3P_t}.
\end{equation}
Here $\Delta\epsilon=\epsilon_{\rm in}-\epsilon_{\rm out}\ge0$ is the outward energy density decrease. The pressure term is retained \cite{SM:takatsy}. At the surface $r=R$, $R$ is the stellar areal radius and $M=m(R)$ the gravitational mass. The surface perturbation determines the dimensionless quadrupolar Love number $k_2$ and dimensionless tidal deformability $\Lambda=2k_2/(3C^5)$ at compactness $C=GM/(Rc^2)$. Identical EOS paths are used for radius, tide and nuclear observables.

\section{Nuclear information bound and tidal response}\label{SM:sec:information}
\subsection{Outcome averaged variance reduction}

Nuclear measurements constrain stellar structure by resolving variation between nuclear inputs. At fixed stellar mass and present data $\mathcal D$, let $X=X(\theta,\phi)$ be a real stellar observable with finite, nonzero posterior variance. Nuclear coordinates are $\theta$. $\phi$ contains core parameters, matching density and stable configuration. All expectations below use the full joint posterior, including correlations induced by physical screening and stellar observations. Integrals over an empirical nuclear prior are weighted sums over its nodes. Define
\begin{align}
 \mu_X(\theta)&=\mathbb E[X\mid\theta,\mathcal D], &
 v_X(\theta)&=\operatorname{Var}(X\mid\theta,\mathcal D),\notag\\
 \overline X&=\mathbb E[X\mid\mathcal D], &
 V_X&=\operatorname{Var}(X\mid\mathcal D)>0.
 \label{SM:eq:conditional_moments}
\end{align}
The conditional mean averages core, matching and configuration uncertainty at fixed nuclear input. Expanding $X-\overline X=(X-\mu_X)+(\mu_X-\overline X)$, the cross term vanishes because $\mathbb E[X-\mu_X\mid\theta,\mathcal D]=0$. Hence
\begin{align}
 V_X&=\mathbb E_{\theta\mid\mathcal D}v_X(\theta)
       +\operatorname{Var}_{\theta\mid\mathcal D}\mu_X(\theta)\notag\\
    &\equiv\mathcal B_{{\rm core},X}+V_{{\rm between},X}.
 \label{SM:eq:general_budget}
\end{align}

A new nuclear only experiment has likelihood $p(y\mid\theta,\mathcal D)$ and satisfies $Y\perp\phi\mid\theta,\mathcal D$. Bayes' rule gives
\begin{align}
 p(\theta,\phi\mid\mathcal D,y)
 &=\frac{p(y\mid\theta,\mathcal D)p(\theta,\phi\mid\mathcal D)}{p(y\mid\mathcal D)}\notag\\
 &=p(\phi\mid\theta,\mathcal D)\,p(\theta\mid\mathcal D,y).
 \label{SM:eq:nuclear_factorization}
\end{align}
For supported outcomes, the conditional distribution of $X$ at fixed $\theta$ therefore retains the moments in Eq.~\eqref{SM:eq:conditional_moments}. Applying total variance after the observation yields
\begin{equation}
 \operatorname{Var}(X\mid\mathcal D,y)
 =\mathbb E_{\theta\mid\mathcal D,y}v_X(\theta)
  +\operatorname{Var}_{\theta\mid\mathcal D,y}\mu_X(\theta).
 \label{SM:eq:updated_budget}
\end{equation}
Let $\mathbb E_{Y\mid\mathcal D}$ average over the predictive density $p(y\mid\mathcal D)=\int p(y\mid\theta,\mathcal D)p(\theta\mid\mathcal D)\,d\theta$. Iterated expectation gives $\mathbb E_{Y\mid\mathcal D}\mathbb E_{\theta\mid\mathcal D,Y}v_X(\theta)=\mathcal B_{{\rm core},X}$. Consequently,
\begin{equation}
 \mathbb E_{Y\mid\mathcal D}\operatorname{Var}(X\mid\mathcal D,Y)
 =\mathcal B_{{\rm core},X}
  +\mathbb E_{Y\mid\mathcal D}\operatorname{Var}_{\theta\mid\mathcal D,Y}\mu_X(\theta)
 \ge\mathcal B_{{\rm core},X}.
 \label{SM:eq:expected_variance_floor}
\end{equation}

The expected reduction also has a positive form that identifies the information learned. Conditional expectation gives $\mathbb E[X\mid\mathcal D,Y]=\mathbb E[\mu_X(\theta)\mid\mathcal D,Y]$. Applying total variance first to $X$ and then to $\mu_X(\theta)$ produces
\begin{align}
 V_X-\mathbb E_{Y\mid\mathcal D}\operatorname{Var}(X\mid\mathcal D,Y)
 &=\operatorname{Var}_{Y\mid\mathcal D}\!\left(\mathbb E[X\mid\mathcal D,Y]\right)\notag\\
 &=\operatorname{Var}_{Y\mid\mathcal D}\!\left(\mathbb E[\mu_X(\theta)\mid\mathcal D,Y]\right)\notag\\
 &=V_{{\rm between},X}
   -\mathbb E_{Y\mid\mathcal D}\operatorname{Var}_{\theta\mid\mathcal D,Y}\mu_X(\theta).
 \label{SM:eq:gain_projection}
\end{align}
Thus the fractional gain obeys
\begin{equation}
 0\le\mathcal T_X
 \equiv1-\frac{\mathbb E_{Y\mid\mathcal D}\operatorname{Var}(X\mid\mathcal D,Y)}{V_X}
 \le\frac{V_{{\rm between},X}}{V_X}
 =1-\frac{\mathcal B_{{\rm core},X}}{V_X}.
 \label{SM:eq:general_ceiling}
\end{equation}
Equality at the upper bound holds precisely when
\begin{equation}
 \operatorname{Var}_{\theta\mid\mathcal D,Y}\mu_X(\theta)=0
 \quad\text{almost surely under }p(Y\mid\mathcal D).
 \label{SM:eq:ceiling_equality}
\end{equation}
The experiment must resolve the nuclear dependent conditional mean of the chosen stellar observable. Exact knowledge of all nuclear coordinates is sufficient, but need not be necessary. A zero gain instead means that the posterior mean of $X$ is unchanged for almost every predictive outcome. These are outcome averaged statements: a particular measured value can increase or decrease the posterior variance. The derivation applies to non Gaussian posteriors and vector valued nuclear measurements without requiring nuclear and core coordinates to be independent in the current posterior.

For squared error prediction, the posterior mean minimizes the expected loss, since any estimate $a$ satisfies
\begin{equation}
 \mathbb E[(X-a)^2\mid\mathcal D,y]
 =\operatorname{Var}(X\mid\mathcal D,y)
  +\big[\mathbb E(X\mid\mathcal D,y)-a\big]^2.
 \label{SM:eq:squared_error_risk}
\end{equation}
The ceiling is therefore the normalized reduction in optimal mean square prediction error provided by perfect nuclear information, evaluated in the stated joint model.

A Gaussian future lead measurement specializes the result to $X=R$. Write $V_R=\operatorname{Var}(R\mid\mathcal D)$ and $Y=p_i+\varepsilon$ at nuclear draw $i$, where $p_i$ is its predicted lead skin and $\varepsilon$ is independent zero mean Gaussian noise with SD $\tau$. Aggregate branch corrected probabilities into nuclear weights $w_i$ and conditional mean radii $\mu_i$, with $\sum_iw_i=1$. For noise SD $\sigma>0$, use the normalized Gaussian density $K_\sigma(t)=(2\pi\sigma^2)^{-1/2}\exp[-t^2/(2\sigma^2)]$. With $\sigma=\tau$, define $\overline R=\sum_iw_i\mu_i$ and $Z(y)=\sum_iw_iK_\tau(y-p_i)$. The updated nuclear weights and centered mean are
\begin{align}
 w_i(y)&=\frac{w_iK_\tau(y-p_i)}{Z(y)},\notag\\
 \mathbb E[R\mid\mathcal D,y]-\overline R
 &=\frac{\sum_iw_i(\mu_i-\overline R)K_\tau(y-p_i)}{Z(y)}.
 \label{SM:eq:lead_posterior_mean}
\end{align}
Substitution into Eq.~\eqref{SM:eq:gain_projection}, with $Z(y)$ as the predictive density, gives
\begin{equation}
 \mathcal T_M(\tau)=\frac{1}{V_R}\int_{-\infty}^{\infty}
 \frac{[\sum_iw_i(\mu_i-\overline R)K_\tau(y-p_i)]^2}{Z(y)}\,dy.
 \label{SM:eq:quad}
\end{equation}
Here $M$ denotes the fixed reference stellar mass, and $p_{\min},p_{\max}$ are the minimum and maximum predicted lead skins among the supported nuclear draws. Quadrature uses 512 Gauss--Legendre points on $[p_{\min}-9\tau,p_{\max}+9\tau]$. This outcome integral is distinct from conditioning on a preferred future central value. The radius gains and core ceilings are evaluated at the same masses. At $1.9M_\odot$, the nuclear only ceiling is 0.9\%.

\subsection{Core prior checks and predictors using radius alone}
High density core assumptions determine how nuclear constraints affect stellar radii and tides. We test this dependence by multiplying existing posterior path weights by alternative core prior factors and normalizing globally, retaining the original likelihoods and physical selection. The alternatives favor $\Delta\epsilon=0$ or 150 MeV fm$^{-3}$ by 4:1, favor $s_2=1$ over $1/3$ by 4:1, restrict to either jump subset, or condition on one matching density. These controlled reweightings isolate the effect of core prior choices on the existing EOS support. Table~\ref{SM:tab:robust} reports the reference mass results.
\begin{table}[ht]\centering
\caption{Reference mass robustness on existing EOS support. Core share, independent 0.030 fm lead gain and cubic held draw log tidal prediction error are percentages. The effective nuclear draw count is $1/\sum_iw_i^2$ for aggregated draw probabilities.}\label{SM:tab:robust}
\begin{tabular}{lrrrr}\toprule
Prior & Core share & Lead gain & Tidal error & Effective draws\\\midrule
Equal grid weights & 70.9 & 14.7 & 30.9 & 192.3\\
Continuous favored 4:1 & 64.3 & 17.7 & 33.5 & 187.8\\
Jump favored 4:1 & 74.7 & 12.8 & 28.0 & 196.7\\
Stiff second segment favored 4:1 & 67.8 & 16.5 & 30.5 & 196.5\\
Continuous only & 58.0 & 20.5 & 34.6 & 184.8\\
Jump only & 75.9 & 12.1 & 26.0 & 199.5\\
$n_m=1.1n_0$ & 32.7 & 28.4 & 29.4 & 152.2\\
$n_m=1.25n_0$ & 44.2 & 27.8 & 27.2 & 195.1\\
$n_m=1.5n_0$ & 79.3 & 10.5 & 31.1 & 191.3\\
\bottomrule\end{tabular}\end{table}

Branch counting preserves each EOS path's total probability at each mass. Let $w_e$ be the current probability of EOS path $e$. If that path has $B_e(M)$ stable configurations supporting that mass, each receives $w_e/B_e(M)$ before normalization over the supporting ensemble. Paths with $B_e(M)=0$ contribute no configuration at that mass. Incomplete support is retained in the probability accounting. The mass grid has 46 points over $1$ to $1.9M_\odot$. The variance identity is checked separately for every prior and mass.

Errors in predicting tides from radius alone quantify the performance of finite predictor classes. At fixed mass, write $x=\ln(R/1\,\mathrm{km})$, $z=\ln k_2$ and $\ell=\ln\Lambda=5x+z+\mathrm{constant}$. Write $\operatorname{Cov}$ for covariance under the current stellar posterior. Minimizing the weighted affine error in $\ell-A-\alpha x$ gives $\alpha_*=5+\operatorname{Cov}(x,z)/\operatorname{Var}(x)$ and $V_{\rm rem}=\operatorname{Var}(z)-\operatorname{Cov}(x,z)^2/\operatorname{Var}(x)$, with $A$ the fitted intercept. The shorthand $\ln R$ used below means $\ln(R/1\,\mathrm{km})$. The resulting residual variance quantifies the information left unresolved by the specified predictor class. The log tidal variance identity is
\begin{equation}
 \operatorname{Var}(\ell)=25\operatorname{Var}(x)+\operatorname{Var}(z)+10\operatorname{Cov}(x,z).
\end{equation}
Its baseline reference mass contributions are 49.2\%, 32.8\% and 18.0\%, summing to 100\%. The covariance contribution carries its algebraic sign in this exact variance budget.

Held draw validation removes all cores and matching choices associated with one nuclear parameter draw before fitting. Cubic predictors use the design $X=(1,t,t^2,t^3)$ for $t=[\ln(R/1\,\mathrm{km})-\ln13]/0.1$. Weighted least squares uses the training moments only. The validation statistic is the sum of weighted squared held draw residuals normalized by the full posterior log tidal variance. Baseline errors are 31.4\% for an affine log radius law, 30.9\% for quadratic and cubic laws, and 30.7\% for a cubic spline with fixed interior radius knots 10 to 17 km at 1 km intervals and boundaries 7 and 22 km. The reweighted analyses refit the cubic relation under each prior. These validation errors assess the tested predictor classes. Optimization over all measurable predictors defines the minimum conditional variance.

The selected nearly equal radius pair demonstrates density profile freedom at fixed nuclear input. Both paths share nuclear draw 241. The first has $n_m=1.1n_0,s_1=0.2,s_2=1,\Delta\epsilon=150$ MeV fm$^{-3}$. The second has $n_m=1.25n_0,s_1=0.2,s_2=1/3,\Delta\epsilon=0$. Unrounded radii are 13.0235413379 and 13.0263966829 km. Tides are 585.3573527 and 647.6281343. Their radius difference is 2.8553 m, tidal fractional difference relative to the first is 10.6381\%, and Love numbers are 0.0884827 and 0.0977883. Independent integrations agree in radius and tide at relative differences below $1.5\times10^{-7}$ for these configurations.

\subsection{Conditional compensation between radius and Love number}
Radius and tide can carry different amounts of nuclear information for the same stellar mass. We compare them using the total variance and outcome average arguments for radius $R$, log tide $\ell=\ln\Lambda$ and log Love number $z=\ln k_2$. The ceiling is the between nuclear draw share of the variance of the specified observable, so its value is not invariant under nonlinear changes of that observable. We evaluate radius and log radius gains alongside the log tidal gain. Conditional means and core variances are recomputed using the full joint weights for every observable, mass, matching choice and core prior reweighting.

Radius and Love number can compensate after the core distribution at a nuclear draw is averaged. With $x=\ln(R/1\,\mathrm{km})$, fixed mass gives $\ell=5x+z+\mathrm{constant}$ and hence $\mu_\ell(\theta)=5\mu_x(\theta)+\mu_z(\theta)+\mathrm{constant}$. For $a=x,z,\ell$, write $\mu_a(\theta)=\mathbb E[a|\theta,\mathcal D]$. Subscripts $\theta$ on variances and covariances denote the current nuclear posterior $p(\theta|\mathcal D)$. The nuclear between draw variance is therefore
\begin{equation}
 \operatorname{Var}_\theta\mu_\ell=25\operatorname{Var}_\theta\mu_x+
 \operatorname{Var}_\theta\mu_z+10\operatorname{Cov}_\theta(\mu_x,\mu_z).
\end{equation}
This covariance is between conditional means, not the covariance over all paired stellar configurations. At $1.4M_\odot$, its signed contributions are $14.0842\%,14.3623\%,-27.0697\%$ of the total log tidal variance, leaving a between draw share $1.3768\%$. The correlation of the two conditional means is $-0.951645$. The full paired stellar sample, in contrast, has positive covariance contribution $18.0\%$ in the total log tidal budget. The two budgets answer different questions and use different random variables.

The compensation is posterior and prior dependent. Before the stellar likelihood, on the original physically screened grid, the between draw log tidal share is $5.7305\%$. Continuous favored, jump favored and stiff favored reweightings after stellar data give $1.4265\%,2.2363\%,1.9164\%$. Fixed match values are $18.0700\%,12.0711\%,3.8581\%$. Marginalizing the uncertain matching density allows its conditional weights to change with the nuclear draw. The pooled ceiling need not lie between the ceilings obtained with each match fixed. These percentages should therefore not be described as a calibration independent nuclear limitation on tides. Table~\ref{SM:tab:channels} makes the chosen comparison explicit.
\begin{table}[ht]\centering
\caption{Reference mass nuclear only ceilings and lead specific expected gains, in percent of the specified observable variance. Matching and core rows use the stellar update.}\label{SM:tab:channels}
\begin{tabular}{lrrrrr}\toprule
& \multicolumn{3}{c}{Any nuclear only experiment: ceiling} &\multicolumn{2}{c}{0.030 fm lead gain}\\
Family & $R$ & $\ln\Lambda$ & $\ln k_2$ & $R$ & $\ln\Lambda$\\\midrule
Before stellar data & 33.89 & 5.73 & 32.47 & 20.33 & 3.23\\
Stellar updated, pooled & 29.11 & 1.38 & 43.82 & 14.68 & 0.26\\
Continuous favored 4:1 & 35.70 & 1.43 & 52.78 & 17.69 & 0.41\\
Jump favored 4:1 & 25.26 & 2.24 & 39.07 & 12.80 & 0.33\\
Stiff favored 4:1 & 32.23 & 1.92 & 43.03 & 16.52 & 0.55\\
$n_m=1.1n_0$ & 67.30 & 18.07 & 37.96 & 28.36 & 7.23\\
$n_m=1.25n_0$ & 55.77 & 12.07 & 37.84 & 27.82 & 5.34\\
$n_m=1.5n_0$ & 20.68 & 3.86 & 55.36 & 10.51 & 1.14\\
\bottomrule\end{tabular}\end{table}

\subsubsection{Structural origin of the primary compensation}\label{SM:sec:structural}
Reconstructed equilibrium profiles connect the conditional covariance to the star's mass distribution. We integrate all 6,885 primary EOS paths at $1.4M_\odot$, retaining their 224 nuclear draws and the original joint GW/NICER weights. At fractional radius $q=r/R$, define $F_q=m(qR)/M$, where $m(r)$ is enclosed gravitational mass and $M$ is total mass. Its nuclear conditional mean is $\overline F_q(\theta)=\mathbb E[F_q|\theta,\mathcal D]$, with the same core, matching and configuration averages used for $\mu_x$ and $\mu_z$. The half mass radius $r_{50}$ satisfies $m(r_{50})=M/2$. Central concentration is also described by $\epsilon_c/\overline\epsilon$, where $\epsilon_c$ is central energy density and $\overline\epsilon=3Mc^2/(4\pi R^3)$.

\begin{figure}[ht]\centering
\includegraphics[width=.95\textwidth]{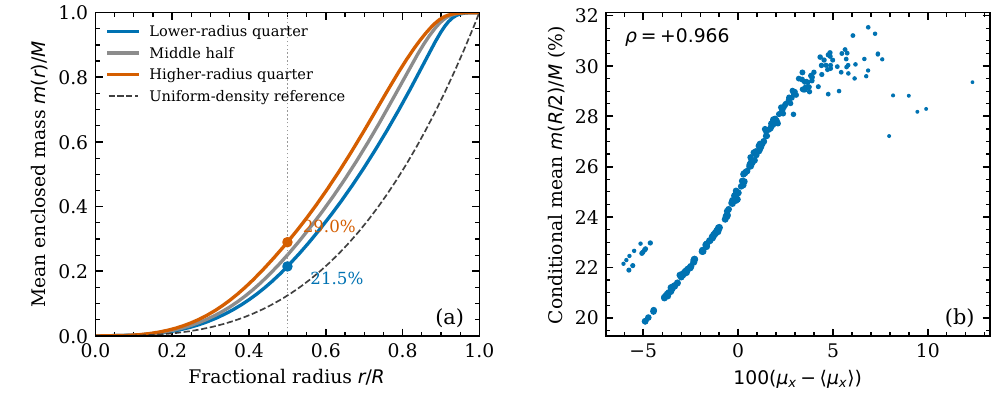}
\caption{Structural counterpart of compensation between radius and Love number in the primary combined posterior at $1.4M_\odot$. (a) Group averaged enclosed mass profiles for nuclear inputs in the lower and higher conditional radius quarters and the middle half. Groups contain 25\%, 25\% and 50\% of nuclear posterior probability. Each nuclear input retains its conditional distribution of cores, matching densities and configurations. The dashed $q^3$ curve is a uniform energy density reference. Dots at $q=1/2$ identify the mean enclosed mass fractions in the two outer groups. (b) Each point gives one nuclear input's conditional mean mass fraction inside $R/2$ against centered mean log radius, with $x=\ln(R/1\,\mathrm{km})$. Marker area follows its posterior probability. The stated correlation uses those probabilities. The curves and points describe equilibrium ensembles rather than a temporal evolution of one star.}\label{SM:fig:structure}
\end{figure}

Figure~\ref{SM:fig:structure} orders nuclear inputs by $\mu_x$ and forms lower and higher quarters containing 25\% of the nuclear posterior probability each, with the middle group containing 50\%. These are weighted equilibrium profile averages. The geometric mean radius increases from 12.211 to 13.086 km between the outer groups, while the mean $F_{1/2}$ increases from 0.2151 to 0.2901 and $\epsilon_c/\overline\epsilon$ from 1.806 to 2.670. The mean half mass radius decreases from 8.399 to 8.241 km, identifying an inward concentration of mass beyond the effect of normalizing the radial coordinate by $R$. Corresponding geometric mean Love numbers are 0.11043 and 0.07797. Geometric mean tides are 529.31 and 528.43. A 7.17\% increase in radius thus accompanies a 29.39\% decrease in $k_2$ and a $-0.17\%$ change in tide. Across all nuclear conditional means, $\overline F_{1/2}$ and $\epsilon_c/\overline\epsilon$ correlate with $\mu_x$ at 0.9656 and 0.9775, respectively. Whole nuclear draw resampling gives 95\% sensitivity ranges 0.9553 to 0.9748 and 0.9686 to 0.9855 for these correlations.

The exact relativistic Love number function $k_2(C,Y)$ separates compactness $C=GM/(Rc^2)$ from the surface tidal response variable $Y=y(R)=RH'(R)/H(R)$, with $H$ the quadrupolar metric perturbation defined above \cite{SM:hinderer}. 

The surface matching expression makes the dependence on compactness and internal response explicit. Put $a(C,Y)=2+2C(Y-1)-Y$ and define
\begin{align}
 d(C,Y)={}&2C\{6-3Y+3C(5Y-8)\}\notag\\
 &+4C^3\{13-11Y+C(3Y-2)+2C^2(1+Y)\}\notag\\
 &+3(1-2C)^2a(C,Y)\ln(1-2C).
 \label{SM:eq:love_denominator}
\end{align}
The relativistic quadrupolar Love number is then
\begin{equation}
 k_2(C,Y)=\frac{8C^5}{5}\,\frac{(1-2C)^2a(C,Y)}{d(C,Y)}.
 \label{SM:eq:love_surface}
\end{equation}
This is the corrected surface matching expression of Ref.~\cite{SM:hinderer}. All internal density jumps enter through the matched value of $Y$, including the junction correction given above. Compactness and $Y$ are evaluated for the same equilibrium solution. The function also defines the mixed reference evaluations used below.

To identify their opposing contributions, write $f(C,Y)=\ln k_2(C,Y)$ and $f_0=f(C_0,Y_0)$, with $C_0=0.1638306$ and $Y_0=0.4467674$ the joint posterior mean values. The symmetric decomposition
\begin{align}
 A_C={}&\tfrac12\{f(C,Y_0)-f_0+f(C,Y)-f(C_0,Y)\},\notag\\
 A_Y={}&f(C,Y)-f_0-A_C
 \label{SM:eq:structure_split}
\end{align}
assigns the nonlinear interaction equally and obeys $f=f_0+A_C+A_Y$ on every reconstructed path. Nuclear conditional averages of $A_C$ and $A_Y$ have weighted regression slopes $+1.064$ and $-5.869$ against $\mu_x$. The net slope of $\mu_z$ is therefore $-4.805$, close to the $-5$ value that offsets the geometric radius factor. Reduced compactness alone favors a larger $k_2$. The opposing change in the integrated internal tidal response is stronger. Varying $C_0,Y_0$ independently over their posterior 16th, 50th and 84th percentiles gives compactness slopes 0.977 to 1.143 and internal response slopes $-5.949$ to $-5.783$, preserving this interpretation. This split is a reference dependent diagnostic, while the summed response and profile comparisons are directly calculated.

The TOV and Love number integration is split at microscopic/crust and core interfaces and includes the Maxwell tidal junction correction. Relative to saved stellar values, reconstructed radii differ by at most 1.737 m and tides by $1.47\times10^{-5}$. The lower surface pressure cutoff shifts the reported radius slightly while leaving the tide stable. The largest target mass difference is $2.11\times10^{-6}M_\odot$. Doubling the microscopic EOS resolution and tightening integration tolerance on six extreme or high weight paths changes tides by at most $4.17\times10^{-6}$ relatively and $F_{1/2}$ by $1.60\times10^{-8}$. The reconstructed log tidal between draw share is 1.37681\%, consistent with the 1.3768\% baseline. Existing quoted budgets retain their original surface convention.

Representative controls retain every accepted core at five nuclear inputs in each continuous core family. Inputs are chosen nearest cumulative nuclear posterior probabilities 0.1, 0.3, 0.5, 0.7 and 0.9 after ordering by $\mu_x$. Table~\ref{SM:tab:profilecontrols} reports these selected point diagnostics. The UNEDF controls retain a strong positive association between conditional radius and concentration. The selected covariant points show a weaker concentration association and a steeper Love response, separating the primary profile pattern from a universal microscopic relation. Five point slopes characterize these controls rather than replace the full family covariance and amplitude tests above. The representative reconstructions agree with their saved radii within $2.03\times10^{-4}$ relatively and tides within $1.25\times10^{-4}$. 
\begin{table}[ht]\centering
\caption{Representative continuous core profile controls at $1.4M_\odot$, with all accepted cores retained for five selected nuclear draws per family. The slope $b_z$ is the weighted regression slope of $\mu_z$ against $\mu_x$ on the selected draws. Correlation refers to $\overline F_{1/2}$ against $\mu_x$. Coverage is their summed nuclear posterior probability. Selection spans the stated quantiles but is not a random posterior subsample.}\label{SM:tab:profilecontrols}
\begin{ruledtabular}\begin{tabular}{lrrrr}
Family & EOS paths & Coverage (\%) & $b_z$ & Concentration correlation\\\hline
UNEDF0 & 755 & 2.59 & $-3.43$ & 1.000\\
UNEDF1 & 761 & 2.02 & $-2.87$ & 0.998\\
Covariant EDF & 810 & 1.96 & $-9.85$ & 0.315\\
\end{tabular}\end{ruledtabular}
\end{table}

\subsubsection{Orthogonal conditional mean decomposition}
The residual tidal response contains both a mismatch of the opposing conditional shifts and scatter around their anticorrelation. An orthogonal projection separates these contributions. Set $u=5\mu_x$ and $v=\mu_z$, centered with their nuclear posterior weights, and, for $\operatorname{Var}(u)>0$, define $\beta=-\operatorname{Cov}(u,v)/\operatorname{Var}(u)$. The residual $e=v+\beta u$ is orthogonal to $u$, giving
\begin{equation}
 \operatorname{Var}(u+v)=(1-\beta)^2\operatorname{Var}(u)+\operatorname{Var}(e).
 \label{SM:eq:mismatch}
\end{equation}
Exact cancellation requires both $\beta=1$ and zero residual variance. The cancellation fraction $\gamma=-2\operatorname{Cov}(u,v)/[\operatorname{Var}(u)+\operatorname{Var}(v)]$ instead measures the share removed from the two positive terms. $\gamma=1$ denotes exact cancellation. At $1.4M_\odot$ in the primary combined posterior, $\beta=0.9610$ and $\gamma=0.9516$. The regression slope mismatch and orthogonal residual terms contribute 0.0214\% and 1.3554\% of total log tidal variance, respectively, summing to the 1.3768\% nuclear only ceiling. The residual is a conditional mean mismatch across nuclear draws, distinct from within draw core scatter. With $\operatorname{Corr}$ denoting correlation and both variances nonzero, defining $\nu=\sqrt{\operatorname{Var}(v)/\operatorname{Var}(u)}$ and $\rho=\operatorname{Corr}(u,v)$ gives $\beta=-\rho\nu$. The orthogonal projection separates slope mismatch from residual scatter. The equivalent amplitude decomposition is $\operatorname{Var}(u+v)=\operatorname{Var}(u)[(\nu-1)^2+2\nu(1+\rho)]$. Both expressions give the same ceiling.

\section{Alternative microscopic and core families}\label{SM:sec:families}
\subsection{Continuous sound speed construction}
A continuous sound speed allows the pressure gradient to vary across the core, providing a second construction of high density matter. Let $\epsilon_m,P_m$ be the matched microscopic energy density and pressure and $s=dP/d\epsilon=c_s^2/c^2$. The microscopic segment fixes $s_m$ at $\epsilon_m$. The values $u_1,u_2,u_3$ at $2\epsilon_m,4\epsilon_m,8\epsilon_m$ are independently uniform on $[0.05,1]$ before physical screening, and $s$ is linearly interpolated in $\ln\epsilon$ between these knots. Above $8\epsilon_m$ it remains $u_3$. Thus pressure and sound speed are continuous and causal, with no Maxwell energy jump. This finite family supplies an alternative to two constant sound speed segments. It is not a nonparametric EOS inference \cite{SM:legred22,SM:legred24}.

Pressure is integrated exactly on each interpolation interval. On adjacent knots $\epsilon_i,\epsilon_{i+1}$, put $s_i=s(\epsilon_i)$, $P_i=P(\epsilon_i)$ and $B_i=(s_{i+1}-s_i)/\ln(\epsilon_{i+1}/\epsilon_i)$. Thus $s(\epsilon)=s_i+B_i\ln(\epsilon/\epsilon_i)$ on that interval, and
\begin{equation}
 P(\epsilon)-P_i=s_i(\epsilon-\epsilon_i)+B_i[\epsilon\ln(\epsilon/\epsilon_i)-\epsilon+\epsilon_i].
\end{equation}
The zero temperature first law defines the positive baryon density by $d\ln n=d\epsilon/(\epsilon+P)$, anchored to the microscopic $n_m$. This guarantees thermodynamic matching. The EOS entering stellar integration is a monotone cubic interpolation of $\epsilon(P)$, with 900 logarithmic core energy samples plus the exact knot locations, extending to $1000\epsilon_m$. A matched crust and microscopic segment are retained as in the primary calculation.

Each calibration supplies its own nonlinear microscopic matter. The UNEDF0 control uses the same 256 node empirical nuclear prior as the primary ensemble. UNEDF1 uses 256 scrambled Sobol Gaussian draws from its original ten dimensional covariance, with inverse isoscalar effective mass active, saturation binding energy per particle of symmetric nuclear matter $E_{\rm NM}=-15.8$ MeV and its incompressibility $K_{\rm NM}=220$ MeV fixed, and inverse isovector effective mass 1.24983857423227. The conversion from these nuclear matter coordinates to Skyrme coefficients reproduces the printed native UNEDF1 solver baseline to maximum absolute coefficient difference $1.82\times10^{-12}$. Uniform matter and stellar structure are recomputed for every draw. Lead skins use the respective validated local HFB sensitivities, as in the primary sampling, so this is nonlinear stellar propagation rather than a nonlinear HFB campaign on every sampled nucleus.

The sound speed quadrature has 64 shared scrambled Sobol core draws at each of the three matches, independent of the nuclear Gaussian quadrature. Both calibrations use identical core coordinates and physical screens. The pre screening attempted count is $2\times256\times3\times64=98,304$ paths. Stability, positive pressure, a valid upper thermodynamic spinodal and maximum mass at least $2M_\odot$ select the surviving paths. Every matched core of a nuclear draw carries equal initial weight and the final prior is normalized globally, preserving differing acceptance fractions. Table~\ref{SM:tab:continuous} reports the physically screened prior comparison. Separate observational updates below apply the same GW and NICER likelihood conventions used by the primary family.
\begin{table}[ht]\centering
\caption{Prior predictive results with continuous sound speed. Gains are expected variance reductions for a 0.030 fm lead measurement, in percent. Core share refers to radius variance. Paths and nuclear draws are accepted counts.}\label{SM:tab:continuous}
\begin{tabular}{crrrrr}\toprule
$M/M_\odot$ & Radius gain & Log tide gain & Core share & Paths & Draws\\\midrule
\multicolumn{6}{c}{UNEDF0: continuous sound speed}\\\midrule
1.2 & 39.8 & 17.7 & 34.9 & 33346 & 224\\
1.4 & 31.3 & 10.9 & 48.9 & 33346 & 224\\
1.6 & 23.3 & 6.6 & 62.0 & 33346 & 224\\
\midrule
\multicolumn{6}{c}{UNEDF1: continuous sound speed}\\\midrule
1.2 & 6.9 & 1.9 & 72.2 & 38293 & 253\\
1.4 & 4.7 & 1.2 & 81.0 & 38293 & 253\\
1.6 & 3.2 & 0.8 & 87.1 & 38293 & 253\\
\midrule
\end{tabular}\end{table}

\begin{figure}[ht]\centering
\includegraphics[width=.95\textwidth]{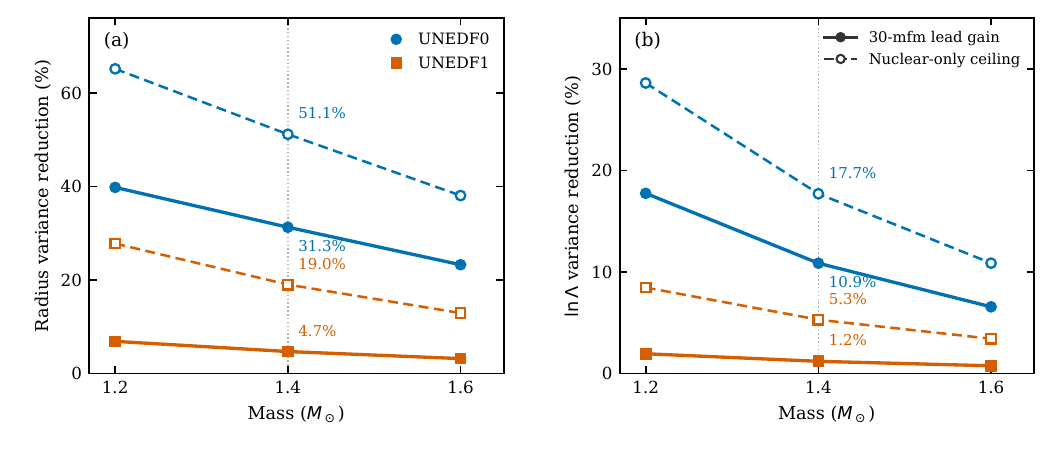}
\caption{Prior predictive nuclear reach in the continuous sound speed families before stellar observations: (a) radius and (b) $\ln\Lambda$ variance reduction. Color and symbol identify UNEDF0 or UNEDF1 consistently in both panels. Solid lines and filled markers give the expected gain from an independent lead measurement with SD 0.030 fm. Dashed lines and open markers give the ceiling for arbitrary nuclear only measurements. Each fraction uses its own calibration and observable variance, with the three matching densities pooled. Labels mark calculated $1.4M_\odot$ values. Straight segments guide the eye between the three calculated masses. The prior includes physical screening and each calibration's local HFB lead response.}\label{SM:fig:continuous}
\end{figure}

An adaptive Dormand--Prince solver integrates mass, radius and tidal perturbation in log pressure using a compiled tabulated EOS. An independent radius coordinate SciPy DOP853 integration checks central pressures 30, 80, 300 and 1000 MeV fm$^{-3}$. The independent integrations differ by less than $4.1\times10^{-8}$ in relative radius and $2.4\times10^{-5}$ in relative tide. Doubling the EOS table resolution or tightening the integration tolerance changes tides by less than $1.7\times10^{-5}$ in six selected controls. Finer maximum mass sampling changes $M_{\max}$ by less than $2.9\times10^{-7}M_\odot$. Sampling checks are given in Table~\ref{SM:tab:sampling}. They separately test EOS/integration error, nuclear quadrature and core quadrature. 
\begin{table}[ht]\centering
\caption{Continuous core sampling controls. Entries are maximum absolute changes over $1.2,1.4,1.6M_\odot$ for the pooled radius results, in percentage points. Nuclear checks compare 128 to 256 draws. Core checks compare 32 to 64 draws.}\label{SM:tab:sampling}
\begin{tabular}{llrr}\toprule
Calibration & Quadrature check & Core share change & Lead gain change\\\midrule
UNEDF0 & Nuclear 128/256 & 0.43 & 0.72\\
UNEDF0 & Core 32/64 & 1.80 & 1.08\\
UNEDF1 & Nuclear 128/256 & 0.09 & 0.08\\
UNEDF1 & Core 32/64 & 1.10 & 0.27\\
\bottomrule\end{tabular}\end{table}

\subsection{Covariant nuclear posterior}
The microscopic calibration sets both neutron skins and the low density pressure entering the stellar inference. We test this connection with the covariant functional calibrated to finite nuclei by Giuliani \etal\ \cite{SM:giuliani23}. We retain the row pairing between their published parameter samples and calculated $^{208}$Pb and $^{48}$Ca neutron skins. The calibration uses ten binding energies and eight charge radii, rather than PREX-II, CREX or stellar likelihoods. Its nuclear posterior therefore acts as a prior for the stellar update here. The 100,000 deposited pairs have a Pb skin mean and SD of 0.2671 and 0.0285 fm. We select 256 rows without replacement using a fixed random seed, preserving the full nonlinear parameter correlations and emulated skin predictions. No UNEDF skin--matter regression is transferred to the covariant ensemble. The forecasts condition on these model skin predictions. An additional neutron skin model discrepancy term is not calibrated here.

The homogeneous matter calculation uses nonlinear scalar, vector and isovector mean fields \cite{SM:chen14}. In natural units $\hbar=c=1$, write $\Phi=g_s\phi$, $W=g_vV_0$ and $B=g_\rho b_0^3$, where $\phi,V_0,b_0^3$ are the scalar isoscalar, vector isoscalar and neutral vector isovector meson fields. Their masses and dimensionless Yukawa couplings define $a_i=m_i^2/g_i^2$ for $i=s,v,\rho$. The field equations are
\begin{align}
 a_s\Phi+\tfrac12\kappa\Phi^2+\tfrac16\lambda\Phi^3&=\rho_s,\\
 a_vW+\tfrac16\zeta W^3+2\Lambda_vWB^2&=n,\\
 (a_\rho+2\Lambda_vW^2)B&=(n_p-n_n)/2.
\end{align}
Here $n=n_p+n_n$ is baryon density and $n_i=k_{Fi}^3/(3\pi^2)$ defines species Fermi momenta for $i=n,p$. The nucleon mass is $m_N$. Each momentum integral is over $|\boldsymbol k|\le k_{Fi}$, with $k=|\boldsymbol k|$. Its factor of two counts spin degeneracy. Scalar density and the effective nucleon mass are
\begin{equation}
 \rho_s=2\sum_{i=n,p}\int^{k_{Fi}}\frac{d^3k}{(2\pi)^3}\frac{m_N^*}{\sqrt{k^2+m_N^{*2}}},\qquad m_N^*=m_N-\Phi.
\end{equation}
The nonlinear couplings $\lambda,\zeta,\Lambda_v$ are dimensionless and $\kappa$ has energy units. Baryonic energy density is
\begin{align}
 \epsilon_b={}&2\sum_{i=n,p}\int^{k_{Fi}}\frac{d^3k}{(2\pi)^3}\sqrt{k^2+m_N^{*2}}
 +\frac{a_s\Phi^2}{2}+\frac{\kappa\Phi^3}{6}+\frac{\lambda\Phi^4}{24}\notag\\
 &+\frac{a_vW^2}{2}+\frac{\zeta W^4}{8}+\frac{a_\rho B^2}{2}+3\Lambda_vW^2B^2.
\end{align}
The baryon chemical potentials are $\mu_{n,p}=\sqrt{k_{Fn,p}^2+m_N^{*2}}+W\mp B/2$, with the upper sign for neutrons and lower sign for protons, and $P_b=\sum_{i=n,p}n_i\mu_i-\epsilon_b$ is baryonic pressure. Free relativistic electrons and muons impose charge neutrality and beta equilibrium. We use $m_N=939$ MeV, $m_v=782.5$ MeV, $m_\rho=763$ MeV and $\hbar c=197.3269804$ MeV fm for unit conversion.

The eight sampled coordinates specify scalar mass $m_s$, saturation Fermi momentum $k_{F0}$, saturation binding energy, effective mass ratio, incompressibility $K$, vector self coupling $\zeta$, saturation symmetry energy $J=S(n_0)$ and slope $L$. The deposited second coordinate, near 1.31 fm$^{-1}$, is converted to $n_0=2k_{F0}^3/(3\pi^2)$. It is not used as a baryon density. Couplings follow the bulk to Lagrangian transformation of Ref.~\cite{SM:chen14}: the saturation pressure condition determines the vector coupling. Energy, scalar stationarity and incompressibility give three linear scalar equations. $J,L$ determine the two isovector couplings. Sixteen sampled points reproduce these bulk constraints numerically. The printed FSUGold2 benchmark gives $J=37.60$ MeV, $L=112.76$ MeV and $K=237.96$ MeV, compared with the published values 37.62, 112.8 and 238.0 MeV.

Crust matching, the thermodynamic spinodal, three microscopic/core matching densities, continuous core sound speed prior, physical screening, source branch averaging and stellar likelihoods follow the same procedures as the UNEDF continuous core controls. This holds the high density construction fixed while changing the nuclear functional and its calibrated uncertainty. Negative or nonmonotonic microscopic segments are excluded by the common screen before stellar reweighting. All 254 surviving nuclear models also pass composition curvature, isovector field and thermodynamic stability checks at twelve densities from just above their spinodal to their largest retained matching density. The covariant screen accepts 40,562 EOS paths. The three mass source and reference roots are calculated directly. Independent DOP853 TOV and Love number integrations agree with the native solver to $2\times10^{-7}$ in relative mass/radius and $1.8\times10^{-5}$ in tide on three representative new paths.

\begin{table}[ht]\centering
\caption{Lead skin widths in the current combined stellar posteriors, before the prospective 0.030 fm measurement. Each SD uses the family's nuclear weights after physical screening and conditioning on GW and NICER data, with matching densities pooled. Published calibration widths and equal node pre screening widths describe different distributions.}\label{SM:tab:skinwidths}
\begin{ruledtabular}\begin{tabular}{lr}
Family & $\sigma_{\Delta_{\rm Pb}|\mathcal D}$ (fm)\\\hline
UNEDF0, two segments & 0.0304\\
UNEDF0, continuous & 0.0311\\
UNEDF1, continuous & 0.0171\\
Covariant, continuous & 0.0275\\
\end{tabular}\end{ruledtabular}\end{table}

\begin{table}[ht]\centering
\caption{Covariant posterior nuclear gains and ceilings. Entries are percentages for a 0.030 fm independent lead measurement. Matching densities are pooled.}\label{SM:tab:covariant}
\begin{ruledtabular}\begin{tabular}{lrrrrr}
Update & $M/M_\odot$ & Gain $R$ & Gain $\ln\Lambda$ & Ceiling $R$ & Ceiling $\ln\Lambda$\\\hline
Before stellar data & 1.2 & 19.13 & 13.84 & 46.50 & 39.02\\
Before stellar data & 1.4 & 11.28 & 7.06 & 27.72 & 21.40\\
Before stellar data & 1.6 & 6.53 & 3.66 & 16.48 & 11.81\\
NICER only & 1.2 & 21.83 & 9.42 & 47.50 & 31.40\\
NICER only & 1.4 & 9.44 & 2.53 & 21.17 & 11.97\\
NICER only & 1.6 & 3.09 & 0.61 & 7.44 & 4.70\\
GW only & 1.2 & 8.56 & 2.02 & 21.89 & 6.57\\
GW only & 1.4 & 2.66 & 0.12 & 7.46 & 0.62\\
GW only & 1.6 & 0.71 & 0.04 & 2.55 & 0.17\\
Combined data & 1.2 & 7.71 & 0.75 & 19.46 & 4.51\\
Combined data & 1.4 & 1.35 & 0.55 & 4.46 & 1.85\\
Combined data & 1.6 & 0.19 & 1.85 & 1.52 & 4.46\\
\end{tabular}\end{ruledtabular}\end{table}

The observational ablations distinguish a conditioned compensation from a universal microscopic relation. Before stellar data, the covariant correlation between conditional mean log radius and log Love number at canonical mass is positive, 0.106. It becomes $-0.959$ after the joint update. The same update gives response amplitude ratio $\nu=1.89$, so strong anticorrelation does not by itself imply the primary posterior's near equality of amplitudes. The nonnegative amplitude/correlation decomposition in the Letter is an exact rearrangement of the covariance budget. It identifies which condition for cancellation fails in a specified posterior, rather than asserting a universal EOS response.

At canonical mass the covariant radius gain has a 95\% whole draw resampling interval 0.43 to 2.71\%, while the log tidal gain has interval 0.26 to 0.95\%. These intervals describe finite cloud sensitivity of small gains. The central values are not precision claims independent of nuclear sampling.

Paired resampling of both observables gives a radius minus log tidal gain interval 0.17 to 1.82 percentage points, using identical draw multiplicities in both budgets. This tests the ordering directly rather than comparing two marginal intervals.

\begin{table}[ht]\centering
\caption{Canonical mass covariant posterior sampling controls. Rows reconstruct and normalize each restricted ensemble. The independent/pool rows use common nuclear inputs and vary only core integration. Entries are percentages.}\label{SM:tab:covariantcontrols}
\begin{ruledtabular}\begin{tabular}{lrrrr}
Rule & Gain $R$ & Gain $\ln\Lambda$ & Ceiling $R$ & Ceiling $\ln\Lambda$\\\hline
256 nuclear / 64 core & 1.35 & 0.55 & 4.46 & 1.85\\
128 nuclear / 64 core & 2.71 & 0.96 & 6.85 & 2.56\\
256 nuclear / 32 core & 2.03 & 0.40 & 5.81 & 2.19\\
128 nuclear / 32 core & 3.57 & 0.81 & 8.38 & 2.77\\
\hline\multicolumn{5}{c}{Independent core rule on common first 128 nuclear draws}\\\hline
Original64 & 2.71 & 0.96 & 6.85 & 2.56\\
Independent64 & 3.55 & 0.71 & 8.48 & 2.07\\
Pooled128 & 3.10 & 0.84 & 7.60 & 2.30\\
\end{tabular}\end{ruledtabular}\end{table}

Nuclear draw resampling and nested rules quantify finite ensemble sensitivity. The independent core scramble holds the first 128 nuclear draws and physical prior ranges fixed, and the pool combines two equally weighted 64 point rules. Posterior, branch and predictive integrations retain their original probabilities. The central result uses the complete 256 by 64 rule. Table~\ref{SM:tab:covariantcontrols} reports the controls, and Table~\ref{SM:tab:covariantforecast} gives the distinct source forecast under the same prospective precisions used for the other three posteriors.

\begin{table}[ht]\centering
\caption{Covariant posterior distinct source forecast at the common reference precisions: radius SD 0.3 km and log binary tidal SD 0.1. The first remaining fraction marginalizes nuclear input. The second holds it exactly known and is normalized by residual core variance. The normalizations are distinct.}\label{SM:tab:covariantforecast}
\begin{ruledtabular}\begin{tabular}{lrr}
Measurements & Remaining total $\ln k_2$ variance (\%) & Remaining core variance (\%)\\\hline
Pulsar radius only & 64.09 & 20.12\\
Radius + binary tide & 31.65 & 8.09\\
\end{tabular}\end{ruledtabular}\end{table}

\begin{table}[ht]\centering
\caption{Covariant forecast sampling sensitivity. Fractions of total current log Love variance remaining, in percent. Nuclear/core restrictions and independent/pool rules use the same physical prior and source likelihood. These differences are finite ensemble sensitivity, not predictive quadrature error or calibrated physical credible intervals.}\label{SM:tab:covariantforecastsampling}
\begin{ruledtabular}\begin{tabular}{lrr}
Rule & Radius alone & Radius + binary tide\\\hline
256/64 & 64.09 & 31.65\\
128/64 & 75.14 & 34.08\\
256/32 & 67.57 & 33.87\\
128/32 & 78.45 & 36.11\\
Independent128/64 & 78.86 & 36.59\\
Pooled128/128 & 76.98 & 35.33\\
\end{tabular}\end{ruledtabular}\end{table}

\section{Observational updates and numerical validation}\label{SM:sec:updates}
\subsection{Source likelihoods and prior corrections}
Astronomical measurements select the nuclear and core combinations consistent with observed radii and tides. The stellar update uses the low spin GWTC-1 GW170817 release \cite{SM:gwtc1}, Riley's J0030 posterior \cite{SM:riley19}, Salmi's updated J0740 posterior \cite{SM:salmi24}, and Miller's J0437 posterior \cite{SM:miller26}. Overlapping alternative analyses are not multiplied. For a released mass and radius density $\widehat p_j(M,R)$, a common source mass density uniform on $[1,3]M_\odot$, stable branch weights $\pi_b(M|E)$ and EOS $E$, the pulse profile likelihood $L_j(E)$ for source $j$ is proportional to
\begin{equation}
 L_j(E)\propto\int_1^3\frac12\sum_b\pi_b(M|E)\omega_j(M)
 \widehat p_j[M,R_{E,b}(M)]\,dM.
\end{equation}
Here $b$ labels a stable branch, $R_{E,b}(M)$ is its radius on EOS $E$, and $\pi_b(M|E)$ is its configuration prior weight under the support convention stated below. The integral measures mass in solar units. The source prior correction factor is $\omega_j=1$ for J0030/J0740 and $\omega_j=M/M_\odot$ for Miller's inverse compactness flat J0437 source. The latter removes the induced $1/M$ prior density in mass and radius coordinates while retaining the radio information once. Unsupported source regions contribute zero. Their weight is not reassigned to a surviving branch.

GW170817's two component tides are evaluated independently on each EOS branch pair. For component gravitational masses $m_1\ge m_2$, chirp mass $\mathcal M=(m_1m_2)^{3/5}/(m_1+m_2)^{1/5}$ and mass ratio $q=m_2/m_1\le1$, component masses are $m_1=\mathcal M(1+q)^{1/5}q^{-3/5}$ and $m_2=qm_1$. The released joint density in $(\mathcal M,q,\Lambda_1,\Lambda_2)$ is integrated using 1,024 common proposals from its mass marginal, dividing by the proposal density in the importance integral. Event mass/spin priors remain in that marginalized likelihood. Component tide source support is $[0,5000]^2$. Stable branches have equal probability at a supported mass and unsupported branch probability is not transferred. Equal weights specify a configuration prior, rather than a population distribution inferred from data. Source prior conventions, not multiplication of overlapping analyses, determine the update.

\subsection{Source separated updates and finite draw sensitivity}
Separate source updates identify which observations select the compensating radius and Love number shifts. We compare the updates on unchanged EOS support. Starting with the globally normalized screened path prior, we apply (i) no stellar likelihood, (ii) the three NICER factors, (iii) GW170817 alone, and (iv) their product. Each case recomputes nuclear draw and conditional core/matching weights. No PREX-II or CREX likelihood enters this baseline before the hypothetical additional lead measurement. Removing a source factor starts from the original prior rather than dividing small posterior probabilities.

The continuous core paths require sequences over source mass support. We reconstruct every already accepted core from its saved nuclear parameters, match and sound speed coordinates. A 96 node logarithmic central pressure grid is augmented by the saved fixed mass pressures and a refined maximum mass point. Shape preserving cubic interpolation represents radius and log tide separately on each increasing mass branch. We retain the original accepted paths and pre screening uniform prior, with global normalization. No new nuclear or core draws enter. The stellar model remains nonrotating, as in the primary calculation.

The NICER integrals reuse the validated density grids, common uniform source mass prior on $[1,3]M_\odot$, explicit prior support boundaries and equal supporting branch probabilities. J0437 retains its mass Jacobian correction. GW170817 uses the same full four dimensional Gaussian kernel density estimate (KDE) and 1,024 mass proposals. Conditional Gaussian mixture algebra evaluates the joint KDE divided by its marginal mass density. Validation compares with the original direct KDE at three tidal scales. Only mass kernels whose squared displacement, contracted with the inverse mass kernel covariance, exceeds 80 are omitted. Unsupported branch probability is never transferred. Numerical checks compare 96 with 192 pressure nodes and 128 with 256 NICER quadrature nodes on selected paths. The half proposal calculation separately tests the effect of using 512 rather than 1,024 GW proposals.
\begin{table}[p]\centering\small\renewcommand{\arraystretch}{1.12}
\caption{Separate stellar updates on each finite family. Gains and ceilings are percentages. $N_{\rm eff,\theta}$ is the effective nuclear draw count.}\label{SM:tab:updated}\begin{ruledtabular}\begin{tabular}{llrrrrrr}
\toprule Update & $M/M_\odot$ & Gain $R$ & Gain $\ln\Lambda$ & Ceiling $R$ & Ceiling $\ln\Lambda$ & $N_{\rm eff,\theta}$ & Paths\\\midrule
\midrule\multicolumn{8}{c}{UNEDF0, two segment cores}\\\midrule
Before stellar data & 1.2 & 31.33 & 7.11 & 52.58 & 12.27 & 221.1 & 6885\\
Before stellar data & 1.4 & 20.33 & 3.23 & 33.89 & 5.73 & 221.1 & 6885\\
Before stellar data & 1.6 & 11.76 & 1.16 & 19.72 & 2.24 & 221.1 & 6885\\
NICER only & 1.2 & 33.71 & 4.83 & 58.55 & 11.61 & 209.6 & 6885\\
NICER only & 1.4 & 17.00 & 1.16 & 29.16 & 4.81 & 209.6 & 6885\\
NICER only & 1.6 & 2.89 & 2.93 & 5.28 & 6.38 & 209.6 & 6885\\
GW only & 1.2 & 27.93 & 9.92 & 55.88 & 20.44 & 179.4 & 6885\\
GW only & 1.4 & 16.89 & 3.45 & 33.53 & 6.98 & 179.4 & 6885\\
GW only & 1.6 & 8.72 & 0.97 & 17.45 & 2.00 & 179.4 & 6885\\
Combined data & 1.2 & 29.50 & 5.59 & 58.34 & 11.31 & 192.3 & 6885\\
Combined data & 1.4 & 14.68 & 0.26 & 29.11 & 1.38 & 192.3 & 6885\\
Combined data & 1.6 & 3.38 & 1.20 & 7.32 & 2.96 & 192.3 & 6885\\
\midrule\multicolumn{8}{c}{UNEDF0, continuous sound speed}\\\midrule
Before stellar data & 1.2 & 39.79 & 17.74 & 65.13 & 28.60 & 221.3 & 33346\\
Before stellar data & 1.4 & 31.27 & 10.86 & 51.14 & 17.70 & 221.3 & 33346\\
Before stellar data & 1.6 & 23.25 & 6.56 & 38.04 & 10.87 & 221.3 & 33346\\
NICER only & 1.2 & 38.30 & 11.70 & 67.01 & 21.12 & 216.2 & 33346\\
NICER only & 1.4 & 28.04 & 3.98 & 48.96 & 7.88 & 216.2 & 33346\\
NICER only & 1.6 & 17.29 & 0.79 & 30.19 & 2.28 & 216.2 & 33346\\
GW only & 1.2 & 29.06 & 7.89 & 54.44 & 14.57 & 192.2 & 33346\\
GW only & 1.4 & 19.69 & 3.14 & 37.24 & 5.97 & 192.2 & 33346\\
GW only & 1.6 & 12.11 & 1.06 & 23.22 & 2.19 & 192.2 & 33346\\
Combined data & 1.2 & 35.29 & 9.48 & 69.00 & 19.25 & 202.4 & 33346\\
Combined data & 1.4 & 25.72 & 2.40 & 50.61 & 5.58 & 202.4 & 33346\\
Combined data & 1.6 & 15.42 & 0.13 & 30.77 & 1.03 & 202.4 & 33346\\
\midrule\multicolumn{8}{c}{UNEDF1, continuous sound speed}\\\midrule
Before stellar data & 1.2 & 6.87 & 1.95 & 27.78 & 8.46 & 252.6 & 38293\\
Before stellar data & 1.4 & 4.70 & 1.20 & 19.01 & 5.30 & 252.6 & 38293\\
Before stellar data & 1.6 & 3.18 & 0.76 & 12.92 & 3.42 & 252.6 & 38293\\
NICER only & 1.2 & 7.04 & 1.01 & 29.19 & 5.42 & 250.5 & 38293\\
NICER only & 1.4 & 4.24 & 0.33 & 17.55 & 2.21 & 250.5 & 38293\\
NICER only & 1.6 & 2.31 & 0.06 & 9.57 & 0.84 & 250.5 & 38293\\
GW only & 1.2 & 5.81 & 1.20 & 25.00 & 5.84 & 243.0 & 38293\\
GW only & 1.4 & 3.54 & 0.56 & 15.37 & 2.85 & 243.0 & 38293\\
GW only & 1.6 & 2.10 & 0.26 & 9.22 & 1.40 & 243.0 & 38293\\
Combined data & 1.2 & 8.71 & 1.48 & 38.25 & 8.30 & 250.2 & 38293\\
Combined data & 1.4 & 5.43 & 0.47 & 23.92 & 3.19 & 250.2 & 38293\\
Combined data & 1.6 & 2.97 & 0.08 & 13.17 & 1.02 & 250.2 & 38293\\
\bottomrule\end{tabular}\end{ruledtabular}\end{table}
\begin{table}[ht]\centering
\caption{Numerical likelihood checks on continuous core paths. Log likelihood differences compare 96/192 pressure nodes and 128/256 NICER quadrature nodes.}\label{SM:tab:newchecks}
\begin{ruledtabular}\begin{tabular}{lrr}
Source & Maximum sequence difference & Maximum quadrature difference\\\hline
J0030 & 4.03e-05 & 2.28e-06\\
J0740 & 9.18e-05 & 5.34e-06\\
J0437 & 5.59e-05 & 0.000904\\
GW & 0.000236 & $\cdots$\\
\end{tabular}\end{ruledtabular}
Checked paths: 87. Conditional mixture/direct KDE maximum log density difference: $3.8e-13$. Doubling GW proposals from 512 to 1,024 changes gains by at most 0.020 percentage points and ceilings by 0.066 percentage points over the tested masses and observables.\end{table}

\begin{figure}[ht]\centering
\includegraphics[width=.87\textwidth]{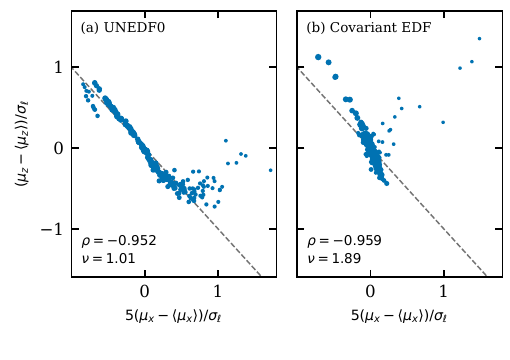}
\caption{Microscopic framework test of conditional compensation at $1.4M_\odot$. The primary UNEDF0 and covariant continuous core posteriors use the same stellar likelihoods. Each point averages over the allowed cores, matches and branches at one nuclear draw. Coordinates are centered on their nuclear posterior averages and normalized by each posterior's total log tidal SD $\sigma_\ell=\sqrt{\operatorname{Var}(\ln\Lambda|\mathcal D)}$. Marker area increases with nuclear draw probability. The dashed line is exact cancellation, not a fitted relation. The two posteriors have similarly strong anticorrelation but different response amplitude ratios $\nu$. Correlation and response amplitude matching determine the strength of cancellation.}\label{SM:fig:mechanism}
\end{figure}

The primary correlations between conditional mean log radius and log Love number at $1.4M_\odot$ are $-0.901$ before data, $-0.872$ for NICER alone, $-0.901$ for GW alone and $-0.952$ jointly. Correlation alone does not determine the ceiling: the amplitudes of the two variance terms matter. The combined update aligns their canceling contributions more closely. For every family and update, direct evaluation of $\operatorname{Var}_\theta(5\mu_{\ln R}+\mu_{\ln k_2})$ agrees with the expanded covariance budget. This mechanism describes how observational conditioning selects the joint radius and Love number response across nuclear inputs. Figure~\ref{SM:fig:mechanism} compares both microscopic frameworks after the joint update.

At $1.4M_\odot$, 2,000 cluster bootstrap replicates resample nuclear draws with replacement, retaining their core, matching and branch states and relative weights. Each replicate recomputes normalized probabilities, variances, ceilings and outcome integrated gains, using the same multiplicities across observables. Table~\ref{SM:tab:bootstrap} gives percentile sensitivity intervals. The scrambled Sobol nodes are not independent random observations, so these intervals describe finite support sensitivity, not calibrated physical confidence or quadrature accuracy. Calibration, core family and likelihood systematic uncertainties are excluded. The effective draw count $1/\sum_iw_i^2$ measures weight concentration.
\begin{table}[ht]\centering\small
\caption{Whole draw resampling sensitivity at $1.4M_\odot$: central values [2.5th, 97.5th percentiles] in percent from 2,000 cluster replicates. These exclude physical model uncertainty.}\label{SM:tab:bootstrap}
\begin{ruledtabular}\begin{tabular}{lrrrr}
Update & Lead gain $R$ & Lead gain $\ln\Lambda$ & Ceiling $R$ & Ceiling $\ln\Lambda$\\\hline
\hline\multicolumn{5}{c}{UNEDF0, two segment cores}\\\hline
Before stellar data & 20.33 [17.13, 23.58] & 3.23 [2.37, 4.24] & 33.89 [30.49, 37.02] & 5.73 [4.67, 6.96]\\
NICER only & 17.00 [14.10, 20.04] & 1.16 [0.77, 1.66] & 29.16 [25.94, 32.33] & 4.81 [3.63, 6.05]\\
GW only & 16.89 [14.26, 19.71] & 3.45 [2.78, 4.22] & 33.53 [30.34, 36.69] & 6.98 [6.02, 8.07]\\
Combined data & 14.68 [12.13, 17.39] & 0.26 [0.15, 0.43] & 29.11 [25.77, 32.37] & 1.38 [1.03, 1.81]\\
\hline\multicolumn{5}{c}{UNEDF0, continuous sound speed}\\\hline
Before stellar data & 31.27 [27.18, 35.37] & 10.86 [8.65, 13.19] & 51.14 [47.37, 54.53] & 17.70 [15.22, 20.19]\\
NICER only & 28.04 [24.26, 31.94] & 3.98 [2.79, 5.45] & 48.96 [45.31, 52.33] & 7.88 [6.08, 9.86]\\
GW only & 19.69 [16.73, 22.61] & 3.14 [2.54, 3.77] & 37.24 [33.93, 40.17] & 5.97 [5.21, 6.73]\\
Combined data & 25.72 [22.44, 28.88] & 2.40 [1.89, 2.96] & 50.61 [47.48, 53.28] & 5.58 [4.87, 6.31]\\
\hline\multicolumn{5}{c}{UNEDF1, continuous sound speed}\\\hline
Before stellar data & 4.70 [3.58, 5.85] & 1.20 [0.87, 1.56] & 19.01 [16.48, 21.31] & 5.30 [4.47, 6.10]\\
NICER only & 4.24 [3.20, 5.29] & 0.33 [0.19, 0.50] & 17.55 [15.22, 19.71] & 2.21 [1.67, 2.80]\\
GW only & 3.54 [2.70, 4.44] & 0.56 [0.42, 0.71] & 15.37 [13.29, 17.31] & 2.85 [2.47, 3.21]\\
Combined data & 5.43 [4.22, 6.63] & 0.47 [0.32, 0.64] & 23.92 [21.16, 26.32] & 3.19 [2.66, 3.75]\\
\end{tabular}\end{ruledtabular}\end{table}

The hierarchy persists for logarithmic radius and tide across all four combined posteriors. The gains in $\ln R$ are 14.38\%, 25.40\%, 5.42\% and 1.35\% for primary UNEDF0, continuous core UNEDF0, continuous core UNEDF1 and the covariant family, respectively. The corresponding ratios of gains in $\ln R$ and $\ln\Lambda$ span 2.4 to 55. These ratios and the absolute gains quantify complementary aspects of experimental reach. In the primary combined posterior at $1.4M_\odot$, the lead gain for $\ln R$ is 14.38\% and its nuclear only ceiling 28.60\%, compared with 0.26\% and 1.38\% for $\ln\Lambda$. The same outcome integral is used for both logarithmic observables.

Experimental thresholds express the same outcome integral as required precision. We solve for the largest independent lead measurement SD yielding a 10\% or 20\% reduction of radius variance. Smaller SD gives at least that gain within this family. Thresholds are checked with doubled integration order. Table~\ref{SM:tab:precision} reports the primary combined posterior results. An unattainable entry exceeds the nuclear only ceiling.
\begin{table}[ht]\centering
\caption{Largest additional lead measurement SD yielding the stated expected radius variance reduction in the primary combined posterior. Precision is in mfm ($10^{-3}$ fm). The ceiling is a percentage.}\label{SM:tab:precision}
\begin{ruledtabular}\begin{tabular}{lrrr}
$M/M_\odot$ & 10\% reduction & 20\% reduction & Nuclear only ceiling\\\hline
1.2 & 66.6 & 42.1 & 58.3\\
1.4 & 42.0 & 20.1 & 29.1\\
1.6 & Unattainable & Unattainable & 7.3\\
\end{tabular}\end{ruledtabular}\end{table}

\subsection{Posterior quadrature and independent core sampling}
Small tidal gains resolve the balance between opposing conditional shifts. We assess their numerical sensitivity directly in the updated posterior. We repeat the combined likelihood calculations on the first 128 versus all 256 nuclear nodes, the first 32 versus all 64 core nodes, and their combined restriction. Each case begins with equal pre screening node weights and recomputes the global posterior normalization, nuclear weights, conditional moments and future measurement outcome integral. Restricting the core rule does not renormalize separately at each nuclear draw. Table~\ref{SM:tab:postquad} reports the resulting changes for radius and log tide at the reference mass. These are quadrature sensitivities rather than confidence intervals or physical changes to the core prior.

\begin{table}[ht]\centering
\caption{Posterior quadrature at $1.4M_\odot$. $N_\theta$ and $N_\phi$ count attempted nuclear nodes and core nodes per matching density. Each restricted rule is screened and normalized globally before applying the same stellar likelihood. Entries are percentages.}\label{SM:tab:postquad}
\begin{ruledtabular}\begin{tabular}{lrrrrr}
Rule & $N_\theta$ & $N_\phi$ & Radius gain & Log tide gain & Log tide ceiling\\\hline
\hline\multicolumn{6}{c}{UNEDF0}\\\hline
Full rule & 256 & 64 & 25.72 & 2.40 & 5.58\\
Nuclear restriction & 128 & 64 & 25.67 & 2.20 & 5.27\\
Core restriction & 256 & 32 & 24.88 & 2.02 & 4.75\\
Both restrictions & 128 & 32 & 24.86 & 1.85 & 4.49\\
\hline\multicolumn{6}{c}{UNEDF1}\\\hline
Full rule & 256 & 64 & 5.43 & 0.47 & 3.19\\
Nuclear restriction & 128 & 64 & 5.50 & 0.46 & 3.04\\
Core restriction & 256 & 32 & 5.11 & 0.40 & 2.82\\
Both restrictions & 128 & 32 & 5.18 & 0.39 & 2.68\\
\end{tabular}\end{ruledtabular}\end{table}

An independent core scramble tests the same continuous sound speed distribution on the common first 128 nuclear nodes. Its 64 three dimensional Sobol points use a distinct scramble, with the same uniform sound speed ranges, matching choices, microscopic construction, physical screens and stellar likelihoods. Globally pooling the two equal size scrambles gives a 128 point core rule on those common nuclear nodes. Table~\ref{SM:tab:independent} compares the original rule, independent rule and their pool. Keeping the nuclear nodes identical isolates the core integration change. Newly accepted paths retain their equal pre screening weight. Unsupported states are not imputed. This control is not a second physical EOS prior, and does not replace the full rule with 256 nodes results used for the main comparison.

\begin{table}[ht]\centering
\caption{Independent core scramble control at $1.4M_\odot$ on the same first 128 nuclear nodes per calibration. The pool combines two equally weighted 64 point core rules before physical screening. Entries are percentages. This is a numerical integration control of the same prior.}\label{SM:tab:independent}
\begin{ruledtabular}\begin{tabular}{lrrr}
Core rule & Radius gain & Log tide gain & Log tide ceiling\\\hline
\hline\multicolumn{4}{c}{UNEDF0}\\\hline
Original 64 & 25.67 & 2.20 & 5.27\\
Independent 64 & 26.35 & 2.45 & 5.77\\
Pooled 128 & 26.01 & 2.32 & 5.51\\
\hline\multicolumn{4}{c}{UNEDF1}\\\hline
Original 64 & 5.50 & 0.46 & 3.04\\
Independent 64 & 5.59 & 0.48 & 3.12\\
Pooled 128 & 5.55 & 0.47 & 3.08\\
\end{tabular}\end{ruledtabular}\end{table}

The independent scramble accepts 35,992 paths from 49,152 attempted combinations. Microscopic exclusions reproduce the baseline stability screens on the same nuclear nodes. Finer sequence and likelihood integration checks on 17 representative independent core paths change log likelihoods by at most 0.00079. These numerical checks are separate from changes induced by resampling core coordinates.

\section{Complementary stellar measurements}\label{SM:sec:forecast}
\subsection{Different source radius and binary tide experiment}
A pulsar radius and the effective tide of a separate binary probe their shared EOS at different stellar masses and density profiles. The reference experiment measures $R(M_R)$ at $M_R=1.418M_\odot$ and $\widetilde\Lambda(\mathcal M,q)$ at $\mathcal M=1.186M_\odot,q=0.85$. The binary component masses are $1.4786578M_\odot$ and $1.2568591M_\odot$. The dimensionless effective tide is $\widetilde\Lambda=16[(1+12q)\Lambda_1+(q+12)q^4\Lambda_2]/[13(1+q)^5]$, where $\Lambda_i$ is the dimensionless tidal deformability of component $i$. This differs physically from the single star Love number benchmark in Sec.~\ref{SM:sec:samestar}.

Independent stable configurations are averaged at a fixed EOS. For EOS path $e$, let $\mathcal R_e$ be supported branches at the pulsar mass and $\mathcal C_{1,e},\mathcal C_{2,e}$ the supported component branches. Let $y_R$ and $y_T$ denote prospective radius and log effective tide measurements. Their independent Gaussian noise SDs are $\sigma_R=0.3$ km and $\sigma_T=0.1$, respectively. Vertical bars on a branch set denote its number of elements. With $R_{e,a}(M_R)$ the pulsar radius on branch $a$ and $\widetilde\Lambda_{e,bc}$ the effective binary tide on component branches $b,c$, the path likelihood is
\begin{align}
 L_e(y_R,y_T)={}&\frac{1}{|\mathcal R_e|}\sum_{a\in\mathcal R_e}K_{\sigma_R}[y_R-R_{e,a}(M_R)]\notag\\
 &\times\frac{1}{|\mathcal C_{1,e}||\mathcal C_{2,e}|}
 \sum_{b\in\mathcal C_{1,e}}\sum_{c\in\mathcal C_{2,e}}
 K_{\sigma_T}[y_T-\ln\widetilde\Lambda_{e,bc}],
\end{align}
where $\sigma_T=\sigma_{\ln\widetilde\Lambda}$. The target $z=\ln k_2$ at $1.4M_\odot$ is also averaged over its own supported branch at fixed EOS, independently of which stars supplied the observations. The predictive variance retains both target moments, with the explicit posterior update derived below. Supported source weight is unity to numerical precision in this case.

The forecast integrates the target posterior variance over the joint predictive observation density. The calculation for radius alone drops the binary likelihood. Source masses are held known at the stated precision pair. The numerical integration uses 256 Gauss--Legendre nodes per observable, checked against 128. One dimensional radius controls use 1024 versus 512 nodes. Domains extend eight noise SDs beyond the predicted extrema. Integrating the predictive density gives unity to better than $10^{-8}$ for the reported cases. A precision grid also varies $\sigma_R=0.1,0.3,0.6$ km and $\sigma_T=0.05,0.1,0.2,0.4$.

Mass uncertainty is a separate linear propagation control. We assign hypothetical independent Gaussian SDs $\sigma_{M_R}=0.03M_\odot$, $\sigma_{\mathcal M}=0.001M_\odot$ and $\sigma_q=0.05$. Derivatives of each calculated stellar sequence add EOS dependent terms to the noise variances,
\begin{align}
 \sigma_{R,e}^2&=\sigma_R^2+[R'_e(M_R)\sigma_{M_R}]^2,\\
 \sigma_{T,e}^2&=\sigma_T^2+
 [\partial_q\ln\widetilde\Lambda_e\,\sigma_q]^2+
 [\partial_{\mathcal M}\ln\widetilde\Lambda_e\,\sigma_{\mathcal M}]^2.
\end{align}
Here $R_e^{\prime}=dR_e/dM$, and partial derivatives of $\ln\widetilde\Lambda_e$ hold the other binary mass coordinate fixed. The effective SDs $\sigma_{R,e},\sigma_{T,e}$ combine measurement noise with the linearized mass uncertainty for each EOS. Central differences use steps 0.001 solar masses, 0.001 and 0.0001 solar masses for these three derivatives, respectively. This is a linearized mass error forecast, not full marginalization over an observed binary mass posterior. A separate stress test introduces experimental radius/log tidal noise correlation $\rho=\pm0.5$ in a bivariate Gaussian at fixed known masses. Here $\rho$ describes measurement noise rather than the conditional mean correlation used in the compensation analysis. Table~\ref{SM:tab:cross} distinguishes these cases.
\begin{table}[ht]\centering
\caption{Different source forecast: remaining log Love number variance, in percent. All cases use 0.3 km radius SD. Joint cases add 0.1 log effective tide SD. Mass and correlated noise rows are separate controls.}\label{SM:tab:cross}
\begin{tabular}{lr}\toprule
Experiment/assumption & Remaining variance\\\midrule
Pulsar radius only & 93.18\\
Radius + binary tide, independent noise & 53.34\\
Radius + binary tide, $\rho=-0.5$ & 57.69\\
Radius + binary tide, $\rho=+0.5$ & 44.13\\
Radius + binary tide, linear mass error & 55.17\\
\bottomrule\end{tabular}\end{table}

The paired source benchmark is repeated under both updated UNEDF continuous core families. The covariant extension is documented above. Exact TOV and Love number roots give the pulsar radius at $1.418M_\odot$ and the two component tides at the stated binary masses. Paths with multiple stable sequence segments are checked on the full sequence with refined turning points. Their supported source configurations receive equal probabilities before likelihood restriction. The target $1.4M_\odot$ branch count is checked against the saved calculation. Table~\ref{SM:tab:crossfamilies} gives both remaining variance fractions and their distinct starting variances.

\begin{table}[ht]\centering
\caption{Cross family paired source forecast at the reference precision pair: pulsar radius SD 0.3 km and binary log effective tide SD 0.1. Remaining fractions refer to each family's current $\ln k_2(1.4M_\odot)$ variance. Starting SDs are dimensionless. Stellar data, source masses and noise conventions are unchanged.}\label{SM:tab:crossfamilies}
\begin{ruledtabular}\begin{tabular}{lrrr}
Updated family & Initial SD & Radius alone (\%) & Radius + tide (\%)\\\hline
UNEDF0, two segment cores & 0.2057 & 93.18 & 53.34\\
UNEDF0, continuous sound speed & 0.1490 & 97.06 & 56.45\\
UNEDF1, continuous sound speed & 0.1048 & 88.28 & 54.07\\
\end{tabular}\end{ruledtabular}\end{table}

Predictive averaging converts the source likelihood into an expected reduction of target uncertainty. Expand each EOS path into its supported source configuration states and label the resulting states by $e$. Their current normalized probabilities $v_e$ include the path probability and the source branch probabilities, so $\sum_ev_e=1$. In this expanded representation, $R_e$ and $\widetilde\Lambda_e$ are the source radius and effective binary tide predicted by state $e$. The reference star's branch is independently averaged at fixed EOS, giving target moments $\overline z_e=\mathbb E[z\mid e,\mathcal D]$ and $\overline{z_e^2}=\mathbb E[z^2\mid e,\mathcal D]$, where $z=\ln k_2(1.4M_\odot)$. The current mean and variance are
\begin{equation}
 \mu_z=\sum_ev_e\overline z_e,\qquad
 V_z=\sum_ev_e\overline{z_e^2}-\mu_z^2.
 \label{SM:eq:target_moments}
\end{equation}
For the random observation vector $Y=(Y_R,Y_T)$ and its realized value $y=(y_R,y_T)$, set
\begin{equation}
 L_e(y)=K_R(y_R-R_e)K_T(y_T-\ln\widetilde\Lambda_e),
 \qquad\int L_e(y)\,dy=1,
 \label{SM:eq:source_state_kernel}
\end{equation}
where $y_T$ is the measured log effective tide, $dy=dy_R\,dy_T$, and $K_R,K_T$ are normalized Gaussian noise densities with the adopted SDs. Independent source noise and conditional independence of the target branch imply that its moments remain fixed within each state $e$. Define the predictive density $D$, centered first moment density $A$ and uncentered second moment density $B_z$ by
\begin{align}
 D(y)&=\sum_ev_eL_e(y),\notag\\
 A(y)&=\sum_ev_e(\overline z_e-\mu_z)L_e(y),\notag\\
 B_z(y)&=\sum_ev_e\overline{z_e^2}L_e(y).
 \label{SM:eq:predictive_densities}
\end{align}
Normalization gives $\int D\,dy=1$, $\int A\,dy=0$ and $\int B_z\,dy=V_z+\mu_z^2$. Bayes' rule assigns state probability $v_e(y)=v_eL_e(y)/D(y)$, and therefore
\begin{align}
 \mathbb E[z\mid\mathcal D,y]&=\mu_z+\frac{A(y)}{D(y)},\notag\\
 \mathbb E[z^2\mid\mathcal D,y]&=\frac{B_z(y)}{D(y)},\notag\\
 \operatorname{Var}(z\mid\mathcal D,y)
 &=\frac{B_z(y)}{D(y)}-\left[\mu_z+\frac{A(y)}{D(y)}\right]^2.
 \label{SM:eq:target_updated_moments}
\end{align}
Multiplying the last line by $D(y)$ and integrating cancels the centered cross term because $\int A\,dy=0$. This gives
\begin{align}
 \mathbb E_{Y\mid\mathcal D}\operatorname{Var}(z\mid\mathcal D,Y)
 &=\int B_z\,dy-\mu_z^2\int D\,dy-2\mu_z\int A\,dy
   -\int\frac{A^2}{D}\,dy\notag\\
 &=V_z-\int\frac{A(y)^2}{D(y)}\,dy.
 \label{SM:eq:predictive_variance_identity}
\end{align}
The same information term is $\operatorname{Var}_{Y\mid\mathcal D}\{\mathbb E[z\mid\mathcal D,Y]\}$. Dividing by $V_z>0$, the expected remaining fraction is
\begin{equation}
 F=1-\frac{1}{V_z}\iint
 \frac{A(y_R,y_T)^2}{D(y_R,y_T)}\,dy_R\,dy_T.
 \label{SM:eq:paired_remaining}
\end{equation}
On any set where $D=0$, the integrand is assigned zero. The identity ensures $0\le F\le1$ and includes unresolved target branch variation through $\overline{z_e^2}$, rather than replacing this second moment by $\overline z_e^{\,2}$. More explicitly, conditional independence and total variance give the lower bound from unresolved target branches
\begin{equation}
 F\ge\frac{\sum_ev_e\left(\overline{z_e^2}-\overline z_e^{\,2}\right)}{V_z}.
 \label{SM:eq:target_branch_floor}
\end{equation}
All forecast masses have single configuration support in the calculated families, as checked in Sec.~\ref{SM:sec:systematics}. The formula also retains branch variance when several target configurations are supported. Forecasting with radius alone drops the binary likelihood, redefines $D,A,B_z$ with the radius kernel alone and evaluates the corresponding one dimensional integral.

Matrix products of one dimensional Gaussian kernel arrays evaluate the direct two dimensional integral without explicitly constructing a likelihood array for every outcome/state pair. The three UNEDF families use 128/256 node Gauss--Legendre checks on domains extending eight noise SDs beyond the source extrema.

An independent Gaussian convolution calculation deposits probability and centered first moment weights onto grids with 20, 40 and 80 bins per noise SD using linear cloud in cell weights, then applies the Gaussian kernels in Fourier space. Direct quadrature and the finest convolution grid are compared for each family and experiment. Far tail cells below $10^{-13}$ of the maximum density are omitted only in the convolution check. Their probability and a bound on omitted information are recorded. Table~\ref{SM:tab:forecastchecks} reports the controls. The reported forecast values use direct quadrature.

\begin{table}[ht]\centering
\caption{Forecast integration checks: maximum absolute changes in percentage points across radius only and paired experiments. Convolution uses 80 bins per noise SD in the final comparison.}\label{SM:tab:forecastchecks}
\begin{ruledtabular}\begin{tabular}{lrr}
Updated family & Direct 128/256 nodes & Convolution/direct\\\hline
UNEDF0, two segment cores & 3.32e-08 & 0.000384\\
UNEDF0, continuous sound speed & 1.28e-12 & 0.000623\\
UNEDF1, continuous sound speed & 4.33e-13 & 0.000337\\
\end{tabular}\end{ruledtabular}\end{table}

\begin{figure}[ht]\centering
\includegraphics[width=.55\textwidth]{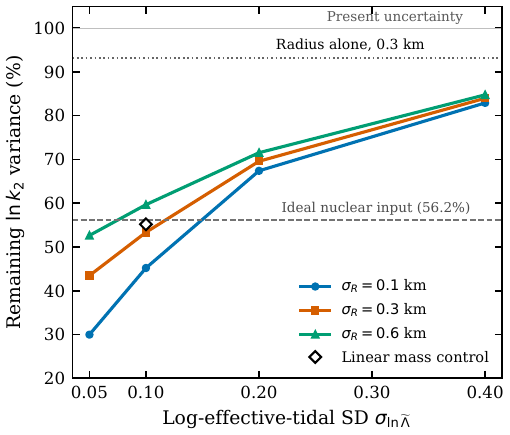}
\caption{Primary posterior precision dependence of the paired source forecast. Curves give expected remaining $\ln k_2(1.4M_\odot)$ variance for three pulsar radius precisions versus log effective tidal SD. Dots are calculated cases joined as guides. Reference levels show present variance, radius alone at 0.3 km SD and exact nuclear parameters. The open diamond adds the linear mass error control at the reference precision pair.}\label{SM:fig:precision}
\end{figure}

We map the precision dependence of the paired source experiment to guide measurement design. Figure~\ref{SM:fig:precision} varies both observational SDs while holding source masses fixed. Its ideal nuclear input reference quantifies the residual uncertainty after exact nuclear conditioning. Additional stellar observations probe that residual directly.

Conditioning on exact nuclear coordinates isolates information carried by the core itself. For experiment $\mathcal E$ (radius alone or the paired sources), define
\begin{equation}
F_{\rm core}(\mathcal E)=
\frac{\mathbb E_{\theta|\mathcal D}\mathbb E_{Y|\theta,\mathcal D,\mathcal E}\operatorname{Var}(z|\theta,\mathcal D,Y,\mathcal E)}{\mathbb E_{\theta|\mathcal D}\operatorname{Var}(z|\theta,\mathcal D)}.
\end{equation}
The denominator is the residual core variance after exact nuclear conditioning. We integrate over outcomes within each nuclear draw and average the resulting variance with its current probability, retaining the same source masses and measurement SDs. This control holds nuclear coordinates known, while the preceding forecasts marginalize them. Table~\ref{SM:tab:coreaccess} shows that stellar measurements resolve the core component itself. The three UNEDF posteriors leave 14.6 to 19.3\% of this residual variance with paired observations, versus 35.4 to 38.0\% for radius alone. The covariant control leaves 8.1\% and 20.1\%, respectively. Direct 64/128 node checks bound numerical integration sensitivity.

\begin{table}[ht]\centering
\caption{Controlled core information test with exact nuclear coordinates. Remaining fractions are percentages of the family's residual core variance in $\ln k_2$, averaged over the current nuclear posterior and predictive measurement outcomes. The last column uses the paired experiment. Nuclear coordinates are held known in this control.}\label{SM:tab:coreaccess}
\begin{ruledtabular}\begin{tabular}{lrr}
Updated family & Radius alone & Radius + tide\\\hline
UNEDF0, two segment cores & 35.40 & 19.34\\
UNEDF0, continuous sound speed & 35.55 & 14.62\\
UNEDF1, continuous sound speed & 37.97 & 15.19\\
\end{tabular}\end{ruledtabular}\end{table}

\subsection{Controlled same star benchmark}\label{SM:sec:samestar}
A paired radius and tide measurement can resolve profile freedom that radius alone poorly constrains. The forecast fixes the true mass at $1.4M_\odot$ and assumes both observables refer to the same star. This controlled experiment isolates same star complementarity. The different source calculation above evaluates a pulsar radius and a separate binary tide through their shared EOS. For branch state $e$ with posterior probability $v_e$, define $r_e=R_e$, $\ell_e=\ln\Lambda_e$ and $z_e=\ln k_{2,e}$. For measured radius $y_R$ and measured log tide $y_\ell$, independent measurement likelihoods are
\begin{equation}
 L_e(y_R,y_\ell)=K_{0.3\,\mathrm{km}}(y_R-r_e)K_{0.1}(y_\ell-\ell_e).
\end{equation}
Forecasting with radius alone drops the second factor. Posterior weights are $v_e(y)=v_eL_e(y)/\sum_fv_fL_f(y)$. For outcome $y$, calculate
\begin{equation}
 V_z(y)=\sum_ev_e(y)z_e^2-[\sum_ev_e(y)z_e]^2.
\end{equation}
The reported remaining fraction is $\mathcal F=\mathbb E_yV_z(y)/\operatorname{Var}(z|\mathcal D)$, averaged over the predictive outcome distribution $p(y|\mathcal D)=\sum_ev_eL_e(y)$. Numerical evaluation may use $z_e=\ell_e-5\ln(r_e/1\,\mathrm{km})$, because its omitted known mass constant cancels in variances.

An independent scrambled Sobol check samples latent states by inverse cumulative posterior probability and Gaussian noise by inverse normal transformation. Two scrambles each use 8,192 outcomes, with a nested 2,048 outcome control. Across twelve forecast cases, the scrambles differ by at most 0.00266 in remaining fraction (0.266 percentage points), and the largest coarse/fine difference is 0.00428. Final results use deterministic Gauss--Legendre integration with 512 nodes for radius alone and a $256\times256$ grid for radius and log tide, extending eight noise SDs beyond the predicted extrema. A $128\times128$ grid differs by at most $1.98\times10^{-7}$ in remaining fraction. Deterministic and averaged Sobol estimates differ by at most 0.00191. Table~\ref{SM:tab:joint} reports the deterministic results, with EOS sampling and model dependence assessed by the separate ensemble and prior checks.
\begin{table}[ht]\centering
\caption{Expected remaining log Love number variance at known mass $1.4M_\odot$, as percentages of the corresponding current prior's variance. Radius SD is 0.3 km. Added log tidal SD is 0.1.}\label{SM:tab:joint}
\begin{tabular}{lrr}\toprule
Core prior & Radius only & Radius and tide\\\midrule
Equal grid weights & 94.5 & 27.4\\
Continuous favored 4:1 & 98.5 & 34.3\\
Jump favored 4:1 & 90.0 & 22.2\\
Stiff second segment favored 4:1 & 96.2 & 30.0\\
Continuous only & 98.7 & 39.7\\
Jump only & 87.0 & 18.6\\
\bottomrule\end{tabular}\end{table}

The variance normalization is specific to each prior. A remaining fraction of 0.274 corresponds to 27.4\% of that prior's current variance of $\ln k_2$, averaged over predictive outcomes. The experiment's log tidal SD 0.1 corresponds approximately to 10\% fractional tidal precision. Known mass, independent errors, the existing EOS support and all present stellar data are assumptions. Correlated radius/tide errors, mass uncertainty, a different nuclear calibration or a different continuous core model require a new calculation.

\section{Configuration support and radius interpretation}\label{SM:sec:systematics}
\subsection{Configuration support}
Stable branch support determines which stellar configurations contribute to each likelihood. Each EOS in all four families supports exactly one source configuration at each of the three forecast masses and one target configuration at $1.4M_\odot$. Thus alternative source or target branch probabilities introduce no additional ambiguity in this experiment at fixed current EOS weights. Multiple stable segments elsewhere on a sequence can still affect the present data likelihood. Their combined posterior probabilities are 0.00058\%, 3.47\%, 3.69\% and 0.88\% for the primary UNEDF0, continuous UNEDF0, continuous UNEDF1 and covariant families, respectively.

As a direct stress test, we exclude every EOS with multiple stable sequence segments, renormalize the remaining current weights globally and repeat the canonical lead update and paired source experiment. Table~\ref{SM:tab:branchsensitivity} reports the resulting changes. The lead radius and log tidal gains change by less than 0.05 percentage points and the paired remaining fractions by less than 0.19 percentage points. This is an exclusion diagnostic, not a fitted branch population model or a bound over arbitrary configuration priors. The single configuration support at the forecast masses and the small exclusion sensitivity identify the role of configuration weighting in the reported results.

\begin{table}[ht]\centering
\caption{Excluding all EOS paths with multiple stable sequence segments. Entries are changes relative to the full current posterior, in percentage points. Positive gain changes mean stronger lead constraints. The final column concerns the paired source remaining $\ln k_2$ variance fraction.}\label{SM:tab:branchsensitivity}
\begin{ruledtabular}\begin{tabular}{lrrr}
Family & $\Delta$ gain $R$ & $\Delta$ gain $\ln\Lambda$ & $\Delta$ paired remaining\\\hline
UNEDF0 two segment & +0.0010 & +0.0000 & +0.0009\\
UNEDF0 continuous & -0.0481 & -0.0379 & -0.0871\\
UNEDF1 continuous & -0.0196 & -0.0060 & -0.1818\\
Covariant continuous & +0.0153 & -0.0213 & +0.1055\\
\end{tabular}\end{ruledtabular}\end{table}

\subsection{Spin scale and radius model sensitivity}
Spin connects the EOS radius to the measured equatorial radius. The stellar sequences here solve the nonrotating TOV and Love number equations. The reused NICER posteriors already account for rotational effects in pulse profile formation. Approximating their equatorial radii by TOV radii is a separate stellar structure approximation. A useful expansion parameter is $\varepsilon_\Omega=(2\pi f)^2R^3/(GM)$, where $f$ is spin frequency. At a fiducial radius of 13 km, the adopted spins of approximately 205.5, 346.5 and 173.7 Hz give $\varepsilon_\Omega=0.0206,0.0379,0.0139$ at representative masses $1.34,2.073,1.418M_\odot$ for J0030, J0740 and J0437, respectively \cite{SM:riley19,SM:salmi24,SM:miller26}. Across 12 to 14 km these parameters span 0.0162 to 0.0257, 0.0298 to 0.0473 and 0.0109 to 0.0174. These values quantify the rotational expansion scale. An EOS specific rotating calculation supplies the corresponding radius correction.

Published rotating star comparisons put the equatorial radius increase at 174 Hz and $1.4M_\odot$ at up to about 0.2 km for particularly stiff, large radius EOSs \cite{SM:choudhury24}. This is below present radius credible interval widths but reaches two thirds of the prospective 0.3 km SD. A source specific rotating calculation is therefore needed to implement that precision experimentally. The paired source calculation specifies Gaussian precision on the modeled EOS radius at known source masses, providing a benchmark for source specific observational modeling.

An additional Gaussian radius nuisance provides a transparent precision sensitivity test. Marginalizing an independent zero mean nuisance of SD $\sigma_{\rm mod}=0.2$ km replaces the benchmark radius SD by $\sqrt{0.3^2+\sigma_{\rm mod}^2}=0.361$ km. The value 0.2 km is an illustrative modeling uncertainty, not an estimated rotation uncertainty or a substitute for a rotating star solver. Table~\ref{SM:tab:radiusmodelsensitivity} gives the result with the binary precision held fixed. Paired measurements still remove 39 to 67\% of the current $\ln k_2$ variance across the four families. Both the nominal and broadened calculations average over predictive outcomes and use the same present stellar posterior. A known, EOS independent radius offset would only translate the outcome coordinate and would leave these outcome averaged gains unchanged.

\begin{table}[ht]\centering
\caption{Sensitivity to an additional independent Gaussian radius model nuisance. Entries are percentages of current $\ln k_2$ variance remaining after the radius and binary tide are paired. The tidal SD remains 0.1 in log space.}\label{SM:tab:radiusmodelsensitivity}
\begin{ruledtabular}\begin{tabular}{lrr}
Family & Nominal $\sigma_R=0.300$ km & Broadened $\sigma_R=0.361$ km\\\hline
UNEDF0 two segment & 53.34 & 55.21\\
UNEDF0 continuous & 56.45 & 61.31\\
UNEDF1 continuous & 54.07 & 54.95\\
Covariant continuous & 31.65 & 32.54\\
\end{tabular}\end{ruledtabular}\end{table}

\begingroup\renewcommand{\bibfont}{\footnotesize}\setlength{\bibsep}{0pt}\sloppy
\endgroup

\end{document}